\documentclass[12pt]{iopart}

\usepackage{iopams}  
\newcommand{\ket}[1]{\left|{#1}\right\rangle}
\newcommand{\bra}[1]{\left\langle{#1}\right|}
\newcommand{\aver}[1]{\left\langle{#1}\right\rangle}
\newcommand{\expect}[3]{\left\langle{#1}\right|{#2}\left|{#3}\right\rangle}
\newcommand{\inner}[2]{\left\langle{#1}|{#2}\right\rangle}
\usepackage{tikz}
\usepackage{verbatim}
\usepackage{siunitx}
\usetikzlibrary{arrows}
\usepackage{hyperref}
\usepackage{graphicx}
\usepackage{caption}
\usepackage{subcaption}

\begin{document}
\title[Estimation of squeezing properties of multiphoton CS from optical tomograms]{Estimation of squeezing properties of multiphoton coherent states from optical tomograms}

\author{Pradip Laha, S Lakshmibala \& V Balakrishnan}

\address{Indian Institute of Technology Madras, Chennai 600036, India}
\ead{slbala@physics.iitm.ac.in}
\vspace{10pt}

\begin{abstract}
We have examined both single and entangled two-mode multiphoton coherent states and shown how the `Janus-faced' properties between two partner states are mirrored in appropriate tomograms. Entropic squeezing, quadrature squeezing and higher-order squeezing properties  for a wide range of nonclassical states are estimated directly from tomograms.  We have demonstrated how squeezing properties of two-mode entangled states produced at the output port of a quantum beamsplitter are sensitive to the  relative phase  between the reflected and transmitted fields. This feature allows for the possibility of tuning the relative phase to enhance squeezing properties of the state. Finally we have examined the manner in which decoherence affects squeezing and the changes in the optical tomogram  of the state due to interaction with the environment.
\end{abstract}

%

\section{Introduction}
\label{Introduction}
\paragraph{} Quantum states of light are identified  from experimentally obtained values of  appropriate observables through detailed reconstruction procedures.  A relevant set of  observables  which can be measured by homodyne detection is provided by the rotated quadrature operators of the radiation 
field. The measured values constitute a quadrature histogram, or optical tomogram, which is the first step in state reconstruction. The reconstruction program crucially relies on the intimate link between optical tomograms, which can be thought of as the marginal distribution functions of continuous variables (corresponding to the rotated field quadratures), on the one hand, 
 and the  Wigner function corresponding to the state, on the other~\cite{bertrand,vogelrisken}. 
 Such a  connection  opens up  
 the possibility of treating the tomogram as the fundamental object which contains complete information about the state. Several nonclassical states of light such as squeezed light have been identified using optical tomography~\cite{smithey,schiller}.  Optical tomograms  also manifest  qualitative signatures of revivals and fractional revivals of the initial state of a system whose time evolution is governed by a nonlinear Hamiltonian~\cite{rohithpra,saumitran}. Further, they can be used to identify if a bipartite state is entangled. This has  been demonstrated~\cite{rohithjosab}  by examining the state at the output port of a quantum beamsplitter for a specific choice of input states.  Continuous-variable optical quantum-state tomography has been reviewed  in~\cite{lvovsky}. 
 
 While these developments  facilitate the understanding of 
 qualitative aspects of continuous-variable tomograms, it is of interest to examine if {\it quantitative} estimates of nonclassical effects such as quadrature squeezing can be obtained  directly from optical tomograms. A preliminary step in this  programme 
is   the identification of distinctive {\it qualitative} signs of  different nonclassical states in tomograms. Verification of quadrature  and 
 entropic uncertainty relations~\cite{mankoeur}, 
 and of  entropic inequalities~\cite{mankoentropy},  provides consistency checks in determining the extent of squeezing  of a state from the tomogram. For instance, for a single-mode system there is an important  bound on the sum of entropies in the position and momentum quadratures,  as well as an  inequality involving the 
 quadrature variance and the corresponding entropy~\cite{orlowski}. In bipartite systems, too,  bounds on the sums of entropies in position and momentum have been established~\cite{mankoeurophysjb}. It is both relevant and important to explore the possibility of  quantifying nonclassical effects directly from the tomogram,  
  without attempting to reconstruct the  state (or density matrix)  from it,   as the latter 
   involves statistical procedures and is inherently error-prone  at various stages. Exploiting optical  tomograms is an alternative to approaches based on obtaining probability distributions of discrete  random variables  such as squeeze tomography~\cite{mankojphysa},  where the statistics of photon number distributions is examined.
 
 An elegant method exists~\cite{wunsche}  
 for estimating quadrature squeezing in single-mode systems directly from  the optical tomogram.  This procedure can be extended in a straightforward manner to quantify both Hong-Mandel~\cite{hong}  
 and Hillery type~\cite{hillery}  higher-order squeezing and two-mode squeezing. 
 This has been used \cite{saumitran}  to evaluate the squeezing properties,  during time evolution,  
 of  a radiation field propagating through an optical medium with cubic nonlinearities, 
and of  a  Bose-Einstein condensate evolving in a double-well potential. However, a comprehensive investigation of signatures of squeezing mirrored in optical tomograms of cat states, multiphoton coherent states (CS) ~\cite{srinivasan},  isospectral counterparts of coherent states~\cite{sesh}  and two-mode squeezed states has not hitherto  
 been undertaken.  In this paper we identify the qualitative differences between tomograms of representative  states of the radiation field, and quantify the extent of entropic and quadrature squeezing of these states directly from tomograms. We investigate a variety  of two-mode  candidate states, including entangled states created at the output port of a beamsplitter.  This investigation is both important and relevant  as cat states can be generated  in practice, and methods have been proposed~\cite{nielsen}  
 to produce superposed large-amplitude 
 coherent states  from two small-amplitude coherent states.
   Further,  `breeding' cat states  by iteratively increasing their numbers experimentally, by means of an ingenious  use of beamsplitters,  has been reported~\cite{etesse}.  Cat states  and multiphoton coherent states (eigenstates of powers of the photon annihilation operator $a$) are ideal candidates for the  investigation at hand, 
 as they display interesting squeezing properties and sub-Poissonian statistics. These states  can be  broadly categorised as annihilation operator eigenstates or  Perelomov-type coherent states~\cite{perelomov}. 
 
 Our choice of states for investigation utilises  a  novel approach to multiphoton coherent states~\cite{srinivasan}  that  highlights relations
  between specific  pairs of  nonclassical states. One starts with the standard commutation relation 
  $[a, a^{\dagger}] = 1$ on  the Hilbert space spanned by the Fock basis $\{\ket{n}, n = 0,1,...\}$ 
 of eigenstates of $a^{\dagger}a$. Define the operators   
 \begin{equation}
 I_{a} = (1+ a^{\dagger}a)^{-1}, \; \;
 G^{\dagger}_{0} = \textstyle{\frac{1}{2}} a^{\dagger\,2} I_{a}, \; \; 
 G^{\dagger}_{1} = \textstyle{\frac{1}{2}} a^{\dagger} I_{a}  a^{\dagger}.
 \label{eqn:IG_0G_1}
 \end{equation} 
  Then, on the `even' subspace  spanned by 
 the set $\{\ket{2n}\}$, we have the commutation relation $[a^{2}, G_{0}^{\dagger}] = 1$. 
 Similarly, on the `odd' subspace spanned by the set $\{\ket{2n+1} \}$, the commutation relation $[a^{2}, 
 G_{1}^{\dagger}] = 1$ holds good. 
 This provides a natural setting for identifying interesting links between certain  sets of states. For instance, eigenstates of $a^{2}$, namely, 
\begin{equation}
\ket{f}_{0} = \exp\,\big(\textstyle{\frac{1}{2}}f a^{\dagger\,2} I_{a}\big) \ket{0}\;\;\text{and}\;\;  
 \ket{f}_{1} = \exp\,\big(\textstyle{\frac{1}{2}}f a^{\dagger} I_{a} a^{\dagger}\big) \ket{1} \;\;(f \in \mathbb{C}),
 \label{eqn:f_0f_1}
 \end{equation}
can be identified  after 
 appropriate changes of variables as  the even coherent state (ECS)  and  the odd coherent state (OCS), respectively. Similarly, the eigenstates of $G_{0}$ and $G_{1}$ (the hermitian conjugates of  $G_{0}^{\dagger}$ and $G_{1}^{\dagger}$) are 
 \begin{equation}
\ket{g}_{0} = \exp\,\big(g a^{\dagger\,2}\big) \ket{0}\;\;
\text{and}\;\; \ket{g}_{1} = \exp\,\big(g a^{\dagger\,2}\big) \ket{1}\;\;
(g \in \mathbb{C}), 
\label{eqn:g_0g_1} 
\end{equation}
 respectively. These states can be identified after  appropriate parameter changes  as the squeezed vacuum state and the Yuen state  \cite{yuen,clmehtayuen}, respectively. The ECS and the squeezed vacuum  (the pair generated from $\ket{0}$) can then be regarded as `Janus-faced' partners, 
 as can the OCS and the Yuen state  (the pair generated from $\ket{1}$).
The concept of Janus-faced partners also extends to the two-mode case.

 The plan of this paper is as follows. In the next section, we summarise the salient features of single-mode and two-mode optical tomograms and examine the qualitative differences between tomograms of various cat states and multiphoton coherent states including Janus-faced partner states.  In particular, we point out how the Janus-faced nature of states is revealed in their tomograms. In Section \ref{sec:squeezing}, we obtain the entropic squeezing properties and the quadrature squeezing properties of these states from the tomograms. The analysis 
 is then extended to  two-mode Janus-faced pairs, namely, the pair coherent states~\cite{agarwalpcs} and the Caves-Schumaker state~\cite{caves1,caves2}. In Section \ref{sec:beamsplitter}, we consider different entangled states produced by sending appropriate factored product cat states through the input ports of a quantum beamsplitter. This device produces a relative phase between the reflected and transmitted fields. The dependence of the squeezing properties  of the output states on this phase is assessed directly from tomograms. In Section \ref{sec:decoherence}, we investigate the  decoherence of the output states 
  when they are subject to amplitude decay or phase damping. An appendix 
  outlines the crucial steps in 
  the evaluation of  quadrature squeezing properties from tomograms
  using the procedure of Ref. \cite{wunsche}.

\section{Optical tomograms}
\label{tomography}  
 \paragraph{}   For a single-mode radiation field, consider the  family of quadrature operators 
 \begin{equation}
   \mathbb{X}_{\theta} = (a e^{-i \theta} + a^\dagger e^{i \theta})/\sqrt{2},
   \label{eqn:xtheta_op}
 \end{equation}
 where $\theta$ ($0 \leq \theta \leq \pi)$, is the phase of the single mode in the homodyne detection setup. It is evident that for $\theta=0$ and $\frac{1}{2}\pi$,  respectively, we have the two field quadrature operators analogous to position and momentum, respectively.  The eigenkets  of  $\mathbb{X}_{\theta}$ are 
 given by 
 \begin{equation}
  \ket{X_{\theta}, \theta} = \frac{1}{\sqrt{\pi}} \exp\,\big(\!-\textstyle{\frac{1}{2}}X_{\theta}^{2}  
  - \textstyle{\frac{1}{2}} e^{i2\theta}a^{\dagger 2}+ \sqrt{2} e^{i\theta} X_{\theta}a^{\dagger}\big) \ket{0}.
  \label{eqn:ket_x_theta1}
 \end{equation} 
 Further, it can be shown that
 \begin{equation}
  \ket{X_{\theta}, \theta} = e^{i \theta a^{\dagger} a}  \ket{X},
  \label{eqn:ket_x_theta2}
 \end{equation}
 where $\ket{X}$ is an eigenstate of $\mathbb{X}_{0}$. 
  The optical tomogram $\omega(X_{\theta}, \theta)$  corresponding to a density matrix $\rho$ is given by 
 \begin{equation}
  \omega(X_{\theta}, \theta) =  \expect{X_{\theta}, \theta}{\rho}{X_{\theta}, \theta}.
  \label{eqn:w_xtheta_rho}
 \end{equation}
$\omega(X_{\theta}, \theta)$ is non-negative and satisfies
 \begin{equation}
  \int \omega(X_{\theta}, \theta) \,dX_\theta= 1
  \label{eqn:w_norm}
 \end{equation}
  and
 \begin{equation}
  \omega(X_{\theta}, \theta+\pi)  = \omega(-X_{\theta}, \theta).
  \label{eqn:w_pi_symm}
 \end{equation}
 It follows from (\ref{eqn:ket_x_theta2}) that 
 \begin{equation}
  \inner{X_{\theta}, \theta}{n} = \frac{e^{-X_{\theta}^2/2}}{\pi^{1/4}} \frac{ e^{-i n \theta}} 
  {\sqrt{2^{n} n!}} H_{n}(X_{\theta}), 
  \label{eqn:xtheta_n}
 \end{equation}
where $H_{n}$ is the Hermite polynomial of order $n$. 
Then, for a pure state $\ket{\psi} = \sum_{n} c_{n} \ket{n}$,  
(\ref{eqn:w_xtheta_rho}) yields the optical tomogram 
 \begin{equation}
  \omega(X_\theta, \theta) = \frac{e^{-X_{\theta}^2}}{\sqrt{\pi}} \Bigg| \sum_{n=0}^{\infty} c_{n} \frac{ e^{-i n \theta}}{\sqrt{2^{n}\: n!}} H_{n}(X_{\theta})\Bigg|^2.
  \label{eqn:w_cnm}
 \end{equation}

 It is straightforward to extend these results to  bipartite systems with two subsystems ($A$ and $B$, say), such as  two radiation fields, or a single-mode radiation field interacting with an atomic medium modelled by an oscillator. The 
 corresponding quadrature operators are 
  \begin{equation}
 \mathbb{X}_{\theta_1} = \frac{1}{\sqrt{2}}(a e^{-i \theta_1} + a^\dagger e^{i \theta_1}), \qquad \mathbb{X}_{\theta_2} = \frac{1}{\sqrt{2}}(b e^{-i \theta_2} + b^\dagger e^{i \theta_2}).
 \label{eqn:xtheta1_xtheta2_op}
 \end{equation}
 The optical tomogram  is given by 
 \begin{equation}
  \omega(X_{\theta{_1}}, \theta_1, X_{\theta{_2}}, \theta_2) =  \expect{X_1, \theta_1, X_2, \theta_2}{\rho_{AB}}{X_{\theta{_1}}, \theta_1, X_{\theta{_2}}, \theta_2},
  \label{eqn:two_mode_w_xtheta_rho}
 \end{equation}
where $\rho_{AB}$ is the  bipartite density matrix. 
It is easy to see that this tomogram  satisfies the requirements of a probability distribution, and that
 \begin{equation}
  \omega(X_{\theta_1}, \theta_{1}+\pi, X_{\theta_2}, \theta_{2}+\pi)  = \omega(-X_{\theta_1}, \theta_{1}, -X_{\theta_2}, \theta_{2}).
  \label{eqn:two_mode_w_pi_symm}
 \end{equation}
Again, for a pure state 
$\ket{\psi} = \sum_{n}\sum_{m} c_{nm}\ket{n}_{1}\otimes \ket{m}_{2}$ 
of the bipartite system we have    
 \begin{eqnarray}
 \fl \omega(X_{\theta_1}, \theta_{1}, X_{\theta_2}, \theta_{2}) = \frac{e^{-X_{\theta_1}^2 - X_{\theta_2}^2}}{\pi} \Bigg| \sum_{n=0}^{\infty} \sum_{m=0}^{\infty} c_{nm}  \frac{  e^{-i n \theta_1} e^{-i m \theta_2}}{\sqrt{n!m! 2^{(n+m)}}} H_{n}(X_{\theta_1})  H_{m}(X_{\theta_2})\Bigg|^2.
 \label{eqn:two_mode_w_cnm}
 \end{eqnarray}
 Since the optical tomogram is normalised as
 \begin{equation}
 \int \int \omega(X_{\theta_1}, \theta_{1}, X_{\theta_2}, \theta_{2}) dX_{\theta_1} dX_{\theta_2} = 1,
 \label{eqn:two_mode_w_norm}
 \end{equation}
 the  tomographic entropy of the bipartite system for any  specific value of 
 $\theta_{1}$ and $\theta_{2}$ 
 can be defined  \cite{mankoentropy} as  
 \begin{equation}
  S_{AB}(\theta_1,\theta_2) =  - \int \omega(X_{\theta_1}, \theta_1, X_{\theta_2}, \theta_{2}) \,\ln\,[\omega(X_{\theta_1}, \theta_1, X_{\theta_2}, \theta_{2})]\, dX_{\theta_1}dX_{\theta_2}.
  \label{eqn:two_mode_tomo_ent}
 \end{equation}
 The tomographic entropy for the subsystem  $A$ for any  specific value of $\theta_{2}$  (equivalently, for a single-mode system) is
 \begin{equation}
  S_{A}(\theta_1) =  - \int \omega(X_{\theta_1}, \theta_1)\,\ln\,[\omega(X_{\theta_1}, \theta_1)]\, dX_{\theta_1},  
  \label{eqn:single_mode_tomo_ent}
 \end{equation}
 where 
 \begin{equation}
  \omega(X_{\theta_1},\theta_1) = \int  \omega(X_{\theta_1}, \theta_{1}, X_{\theta_2}, 
  \theta_{2}) dX_{\theta_2}.
  \label{eqn:red_w_xtheta}
 \end{equation}
 It is clear that $S_{A}$ is independent of the value of $\theta_{2}$ chosen, and that
 \begin{equation}
  \int \omega(X_{\theta_1}, \theta_1) dX_{\theta_1} = 1.
  \label{eqn:red_w_norm}
 \end{equation}
 A similar definition can be given for the entropy of subsystem $B$. 

 The entropic uncertainty relation for the bipartite system  is given by \cite{orlowski}
 \begin{equation}
  S_{AB}(\theta_1,\theta_2) + S_{AB}(\theta_1+\textstyle{\frac{1}{2}}\pi, 
  \theta_2+\textstyle{\frac{1}{2}}\pi) \geq 2\,\ln\,(\pi e),
  \label{eqn:eur_two_mode}
 \end{equation}
 and correspondingly, for a subsystem ($A$, say)  by
 \begin{equation}
  S_{A}(\theta_1) + S_{A}(\theta_1+\textstyle{\frac{1}{2}}\pi) \geq \ln\, (\pi e).
  \label{eqn:eur_single_mode}
 \end{equation}
 A state with entropy in either quadrature less than $\textstyle{\frac{1}{2}} \ln\,(\pi e)$  is said to display entropic squeezing in that quadrature.

 In order to estimate the extent of quadrature squeezing or higher-order squeezing  of a state from a tomogram, the crucial step  is to compute \cite{wunsche} 
  the expectation value of products of powers of $a$ and $a^{\dagger}$ (see the 
  Appendix). 
  In the single-mode case,   this expression is given by 
 \begin{equation}
 \fl \aver{a^{\dagger\,k}a^{l}} = C_{kl}\sum_{m}^{k+l} \exp\Big(\!-\frac{i(k-l)m\pi}{k+l+1}\Big) \int_{-\infty}^{\infty} dX_{\theta} \: \omega\Big(X_{\theta}, \frac{m\pi}{k+l+1}\Big) H_{k+l}(X_{\theta}),
 \label{eqn:exp_ak_al}
 \end{equation}
 where $C_{kl} = k! \,l!/((k+l+1)!\sqrt{2^{k+l}})$.
 It is straightforward to generalise this result to the two-mode case. 
It  is clear from (\ref{eqn:exp_ak_al})
that we need to consider $(k + l + 1)$ values of the tomogram variable $\theta$ in order to calculate a moment of order $(k + l)$ from a single tomogram.  That is, $(k + l + 1)$ probability distributions $\omega(X_{\theta})$ corresponding to these selected values of $\theta$ are used  to calculate the extent of squeezing and higher-order squeezing, without indulging in an elaborate  state-reconstruction program. 
 In the case of a  system evolving  in time, the extent of squeezing at various instants is determined from the corresponding instantaneous tomograms. 

 \subsection{Tomograms of single-mode states}
 \paragraph{}  We first express the multiphoton coherent states  of interest to us in a manner convenient for numerical computations. In terms of the 
 CS $\ket{\alpha}$  (where $\alpha \in \mathbb{C}$), we have
 \begin{equation}
  \textrm{ECS} = \mathcal{N}_{\alpha_{+}} \big(\ket{\alpha} + \ket{-\alpha} \big), \qquad  \textrm{OCS} = \mathcal{N}_{\alpha_{-}} \big(\ket{\alpha} - \ket{-\alpha} \big)
  \label{eqn:ecsocs}
 \end{equation}
  with normalization constants  $\mathcal{N}_{\alpha_{\pm}} = 
  [2(e^{ |\alpha|^{2}} \pm e^{- |\alpha|^{2}}]^{-1/2}$.
   The  Yurke-Stoler state (YS) is given by
 \begin{equation}
  \textrm{YS} = \big(\ket{\alpha} + i\:\ket{-\alpha}\big)/\sqrt{2}.
  \label{eqn:ys_state}
 \end{equation}
 Writing the complex number $g$ in the form  
 $\xi/|\xi| \tanh|\xi |$ ($\xi \in \mathbb{C}$),  
 the squeezed vacuum state $\ket{g}_{0}$ and the Yuen state  $\ket{g}_{1}$, defined in 
 (\ref{eqn:g_0g_1}), 
 can be expressed as $\mathcal{S}(\xi)\ket{0}$ and $\mathcal{S}(\xi)\ket{1}$, 
  respectively, where 
  \begin{equation}
  \mathcal{S}(\xi) = \exp\,[\textstyle{\frac{1}{2}}(\xi^{*} a^{2} - \xi a^{\dagger^2})]
  \label{eqn:smsqueezeoper}
  \end{equation}
    is the  standard single-mode squeezing operator.  
 
 \begin{figure}
  \centering
  \includegraphics[width=0.244\textwidth,angle=-90]{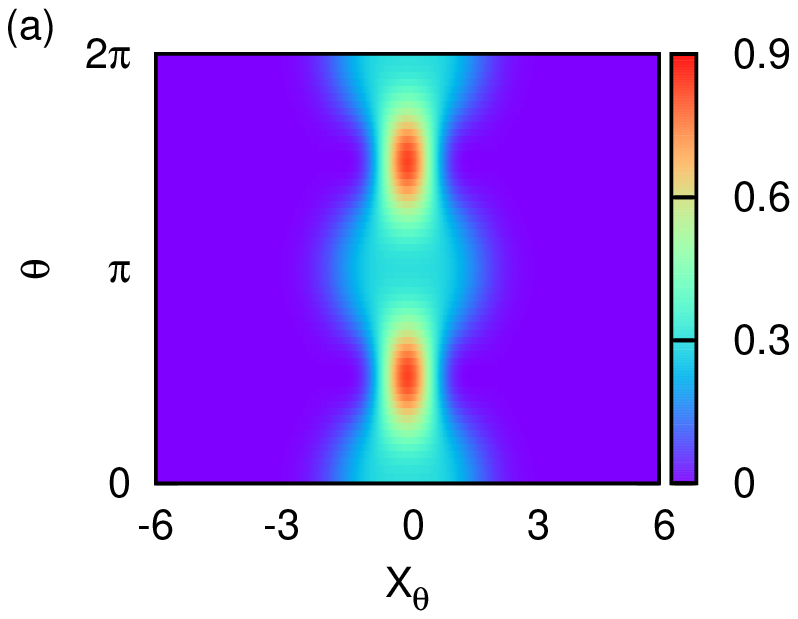}\hspace{0ex}
  \includegraphics[width=0.244\textwidth,angle=-90]{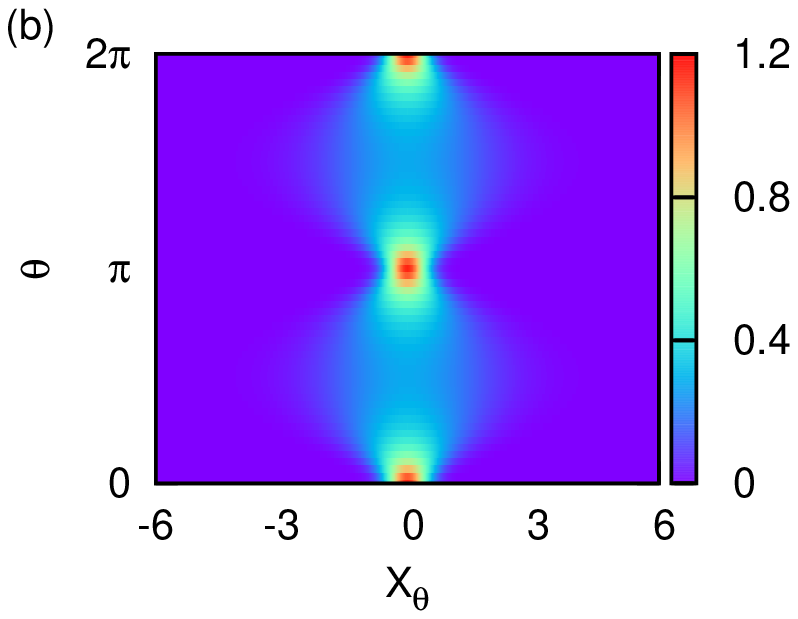}
  \includegraphics[width=0.244\textwidth,angle=-90]{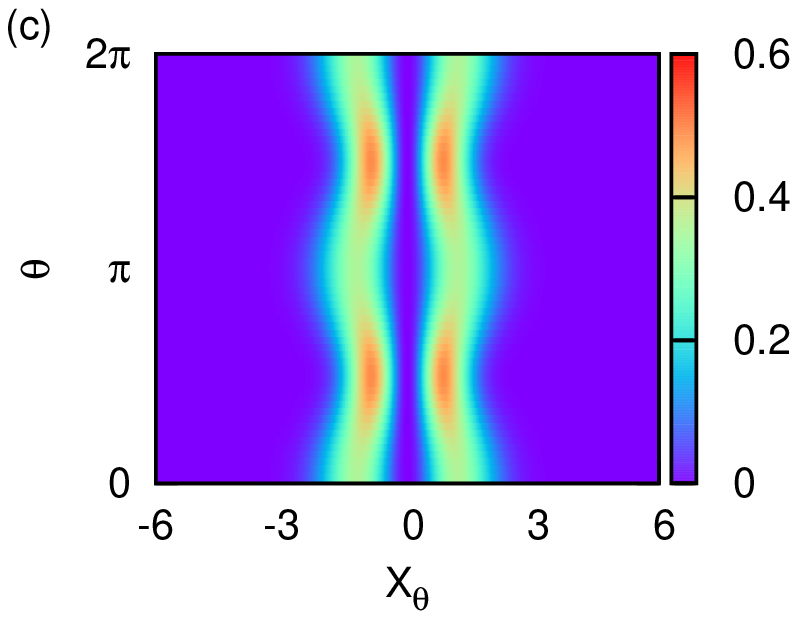}
  \includegraphics[width=0.244\textwidth,angle=-90]{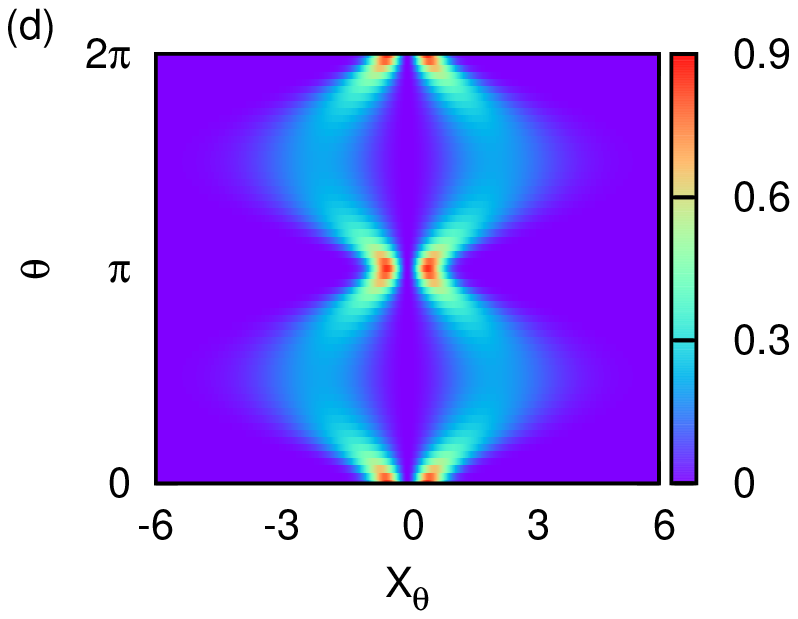}
  \includegraphics[width=0.244\textwidth,angle=-90]{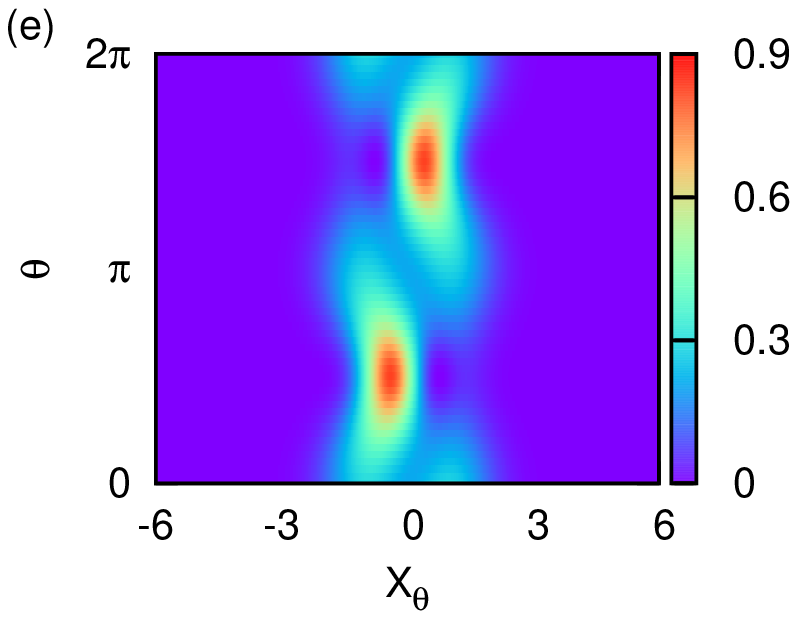}
  \includegraphics[width=0.244\textwidth,angle=-90]{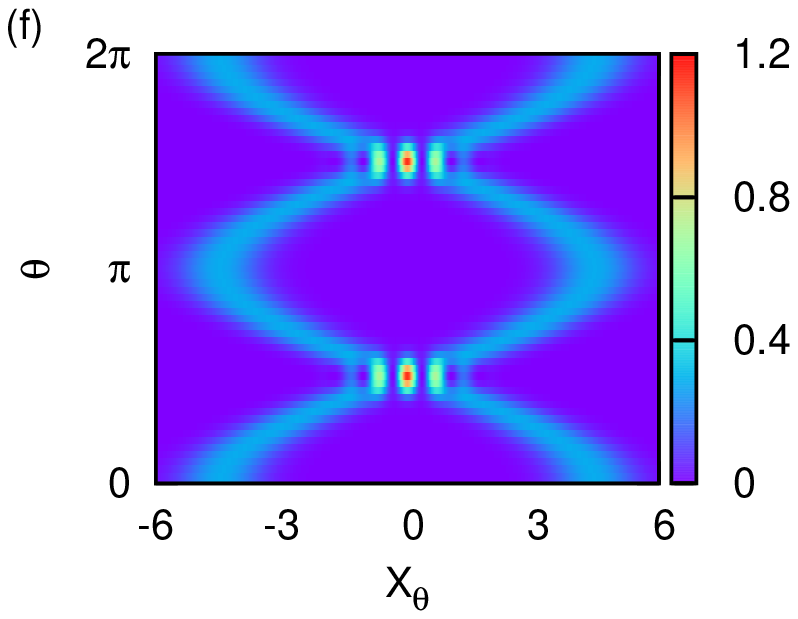}
  \caption{Tomograms for  (a) ECS, (b) squeezed vacuum, (c) OCS, (d) Yuen and  (e) YS states with $\alpha = \xi = 1/\sqrt{2}$ and (f) ECS/OCS/YS state with $\alpha = \sqrt{10}$.}
  \label{fig:singlemode_tomo_catstate}
 \end{figure}
 We have also examined the tomogram corresponding to the 
 isospectral coherent state~\cite{sesh}.  This state differs from 
 the coherent states considered in the foregoing, in the following sense. 
 Consider the operator
 \begin{equation}
  a_{1} =  a^{\dagger}(1+ a^{\dagger}a)^{-1/2} a (1+a^{\dagger}a)^{-1/2} a.
 \label{eqn:icsoperator}
 \end{equation}
It is easily checked that $ [a_{1}, a_{1}^{\dagger}] = 1- \ket{0}\bra{0}$,  so 
that $a_{1}$ annihilates both $\ket{0}$ and $\ket{1}$.  In the restricted Hilbert space with basis $\{\ket{n_{1}}\}$ ($n_{1} = 1,2,\ldots)$, we can therefore define the isospectral coherent state 
   \begin{equation}
   \ket{\zeta, 1} = \exp\,(\zeta a_{1}^{\dagger} - \zeta^{*} a_{1}) \ket{1}, 
 \;\;\zeta \in \mathbb{C}.
 \label{eqn:icsdefn}
 \end{equation}
    This state is an eigenstate of $a_{1}$, with eigenvalue $\zeta$. This idea can be extended to other restricted Hilbert spaces with bases $\{\ket{n_{i}}\}$ 
    ($n_{i} = i, i+1, \ldots$). We show below that there are subtle qualitative differences  between the tomograms of $\ket{\zeta,n_{i}}$ and the tomograms corresponding to the $m$ photon-added  coherent state ($m$-PACS) \cite{tara} $\ket{\alpha, m}$, 
    obtained by normalising the state 
$a^{\dagger\,m}\ket{\alpha}$ to unity. 
$\ket{\alpha, m}$ 
is a {\em nonlinear} coherent state, i.e., it is an eigenstate of 
the operator $[1- m (1+ a^{\dagger} a)^{-1}] \,a$ \cite{sivakumar}.

\begin{figure}
  \centering
  \includegraphics[width=0.235\textwidth,angle=-90]{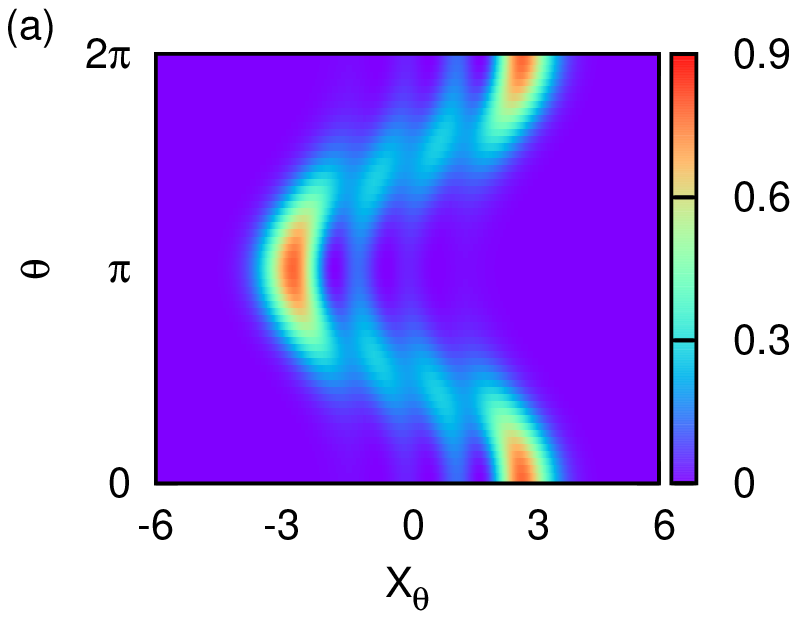}
  \includegraphics[width=0.235\textwidth,angle=-90]{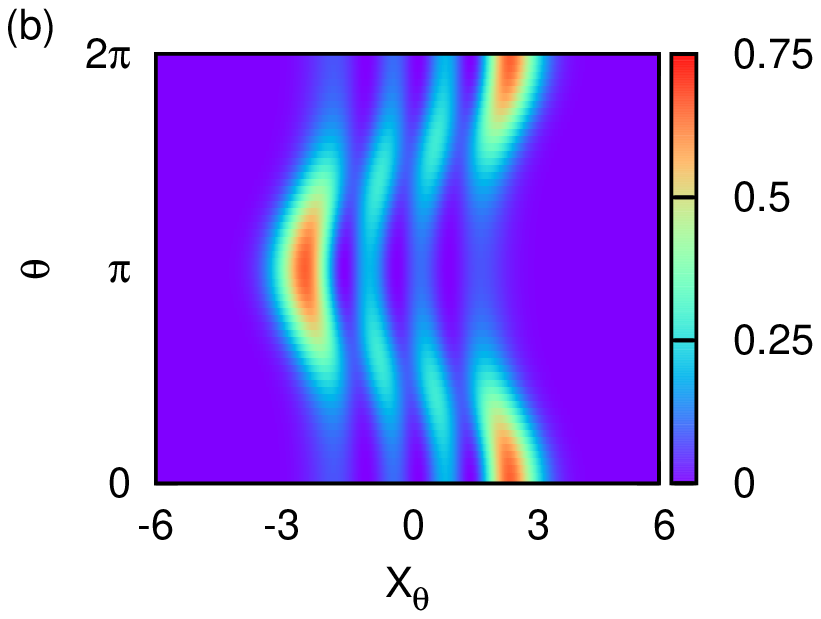}
  \includegraphics[width=0.235\textwidth,angle=-90]{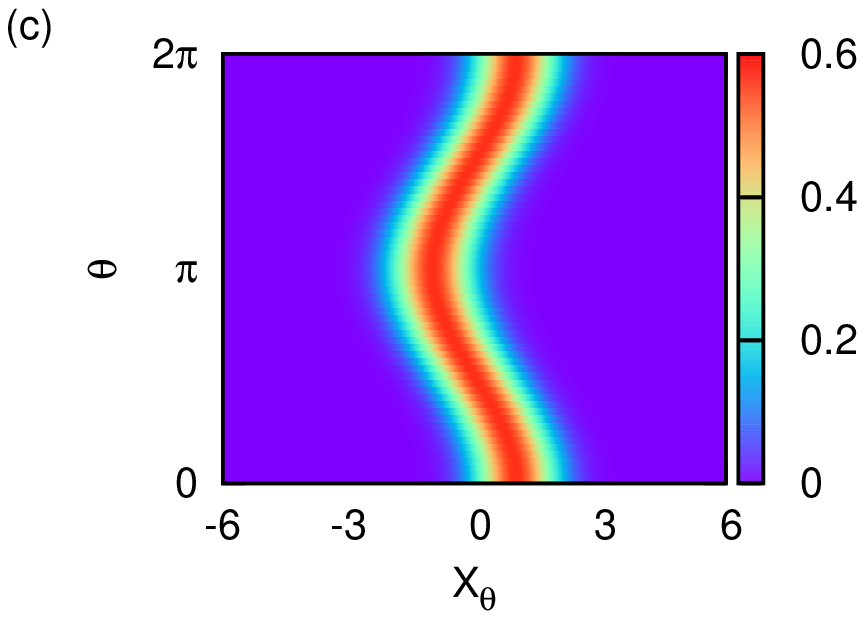}
  \caption{Tomograms  for (a) $\ket{\alpha, 3}$, (b) $\ket{\zeta, 3}$ and (c) $\ket{\alpha}$, with $\alpha = \zeta = 1/\sqrt{2}$.}
  \label{fig:singlemode_tomo_pacs_scs}
 \end{figure}

 Figures \ref{fig:singlemode_tomo_catstate}(a-f) and  \ref{fig:singlemode_tomo_pacs_scs}(a-c)  are the tomograms for the single-mode states defined above. We have taken  the parameters $\alpha$, $\xi$  and $\zeta$ to be real, without significant 
 loss of generality.  As expected, for sufficiently large values of $|\alpha|$,  
 the tomograms for the ECS, OCS and the YS state are identical  (figure \ref{fig:singlemode_tomo_catstate}(f)). The Janus-faced 
 nature  of partner states is revealed in the tomograms. The qualitative appearance (apart from a phase difference of $\textstyle{\frac{1}{2}}\pi$)  of the tomograms for the ECS and the squeezed vacuum state [respectively, the OCS and the Yuen state] are very similar: compare figures \ref{fig:singlemode_tomo_catstate}(a) and  (b)) [resp., figures \ref{fig:singlemode_tomo_catstate}(c) and (d)]. Since the ECS and OCS are superpositions of two coherent states, their tomograms  are two-stranded. It is interesting that this property is also displayed by the tomograms of  their Janus-faced partners.  As expected, this feature holds for the YS state as well,  and   becomes more evident for large values of 
 $|\alpha|$.  

  For sufficiently small vales of $|\alpha|$,  the tomogram of an $m$-PACS has $m$ vertical bands. This feature is present also in the tomogram for the isospectral CS constructed from $\ket{m}$ (figures \ref{fig:singlemode_tomo_pacs_scs}(a) and (b)). The bands are more prominent in the latter tomogram, for the same value of $\alpha$. However, 
  this feature disappears for large 
  $|\alpha|$, and the tomograms corresponding to the isospectral CS and the $m$-PACS are similar to that of the standard CS (figure \ref{fig:singlemode_tomo_pacs_scs}(c)). 

 \subsection{Tomograms of two-mode Janus-faced partners }
 \label{two_mode_janus_face}
\begin{figure}
  \centering
  \includegraphics[width=0.35\textwidth,angle=-90]{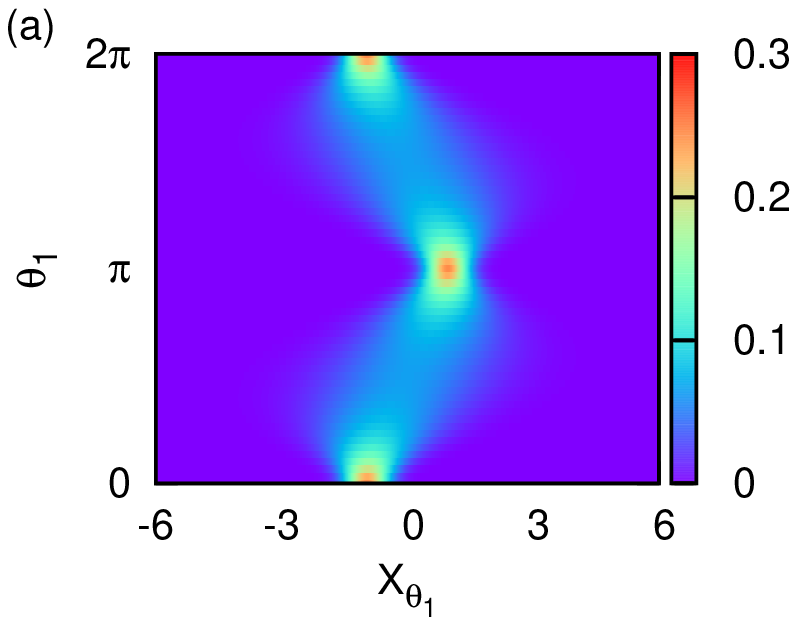}
  \includegraphics[width=0.35\textwidth,angle=-90]{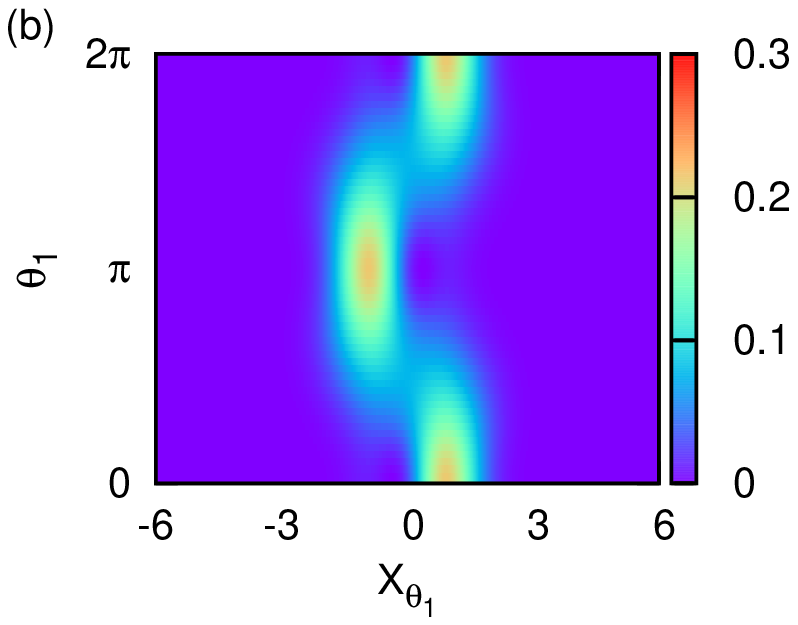}
  \caption{Tomograms for (a) Caves-Schumaker and (b) pair coherent states with $\eta=1$, $X_{\theta_{2}} = 1$ and $\theta_{2} =0$.}
  \label{fig:twomode_tomo}
 \end{figure}

 \paragraph{}  
 We recall that  the  photon destruction and creation operators  for the two modes $A$ and $B$ are  $(a,a^{\dagger})$ and $(b,b^{\dagger})$, respectively. The relevant commutator  in this case is  given by 
 $[a b, G^{\dagger}] = 1$, where   $G^{\dagger} = a^{\dagger} b^{\dagger} (I_{a} + I_{b})$, $I_{a} = (1+ a^{\dagger}a)^{-1}$ and $I_{b} =(1+ b^{\dagger}b)^{-1}$. Let $\ket{0,0}$ denote the direct product ground state 
 $\ket{0}_{A} \otimes \ket{0}_{B}$. The pair coherent state~\cite{agarwalpcs} is given by $\exp\,(f^{\,\prime} G^{\dagger})\ket{0,0}$ ($f^{\,\prime} \in \mathbb{C}$). Its Janus-faced partner is the Caves-Schumaker state $ \exp\,(g^{\,\prime} a^{\dagger} 
 b^{\dagger})\ket{0,0}$ ($g^{\,\prime} \in \mathbb{C}$). By a suitable change of variables we can identify the Caves-Schumaker state  as  the two-mode squeezed state $\mathcal{S}_{AB}(\eta) \ket{0, 0}$ \cite{caves1,caves2}, where the two-mode squeezing operator 
 \begin{equation}
 \mathcal{S}_{AB}(\eta) = \exp\,[\textstyle{\frac{1}{2}}(\eta a^{\dagger}b^{\dagger} - 
 \eta^{*} ab)], \,\eta \in \mathbb{C}.
 \label{eqn:twomodesqueezeoper}
 \end{equation} 
 Setting $\eta = r e^{i\theta}$, the Caves-Schumaker and pair-coherent states 
 can be expressed in the two-mode  Fock basis as 
\begin{equation}
\text{sech}\,r \sum_{n=0}^{\infty} e^{n i\theta} (-\tanh r)^{n} \ket{n, n}\;\;
\text{and}\;\; N_{0} \sum_{n=0}^{\infty} \frac{r^{n} e^{ni\theta}}{n!} \ket{n;n},
\label{eqn:caves-schu-paircoherent}
\end{equation}
respectively. Here  
 $N_{0} = 1/\sqrt{I_{0}(2r)}$, $I_{0}$ denoting the modified 
 Bessel function of order $0$. 
The tomograms  for the two-mode squeezed state and the pair coherent state  (figures \ref{fig:twomode_tomo}(a) and (b)) are qualitatively similar; the only difference is 
due to a phase shift. (For illustrative purposes, we have set 
$\eta = 1$ in both cases.) The intensity of the tomograms depends on 
the specific value of $X_{\theta_{2}}$.

\section{ Estimation of squeezing properties from tomograms}
 \label{sec:squeezing}
\begin{figure}
  \centering
  \includegraphics[scale=0.575]{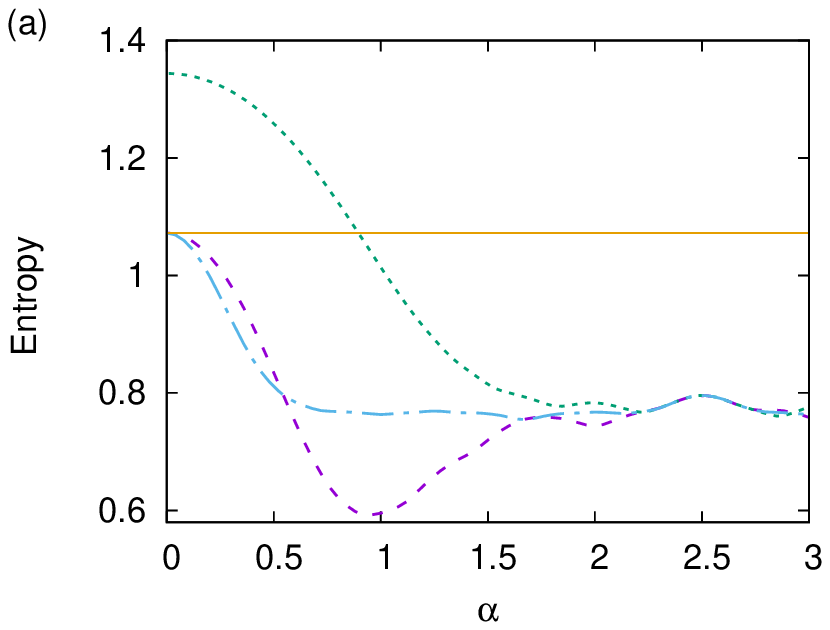}
  \includegraphics[scale=0.575]{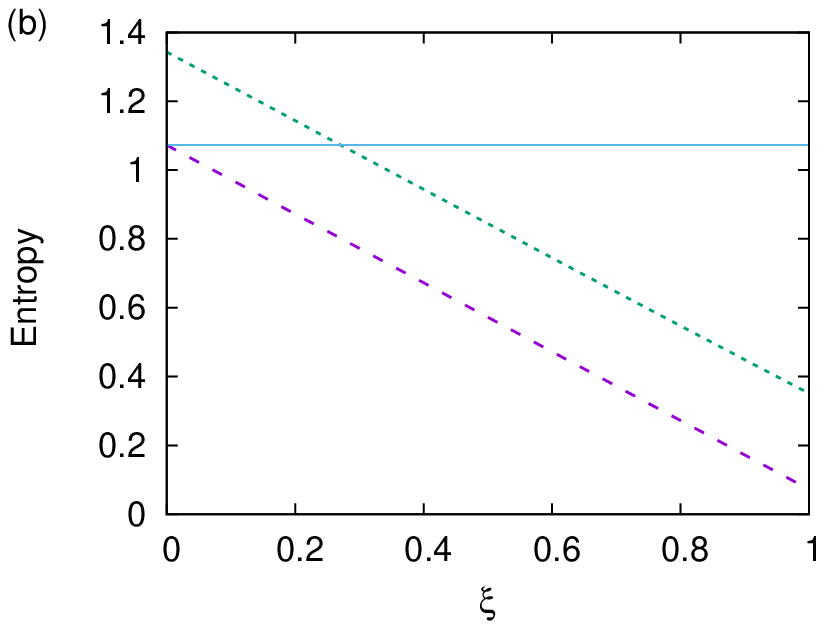}
  \includegraphics[scale=0.575]{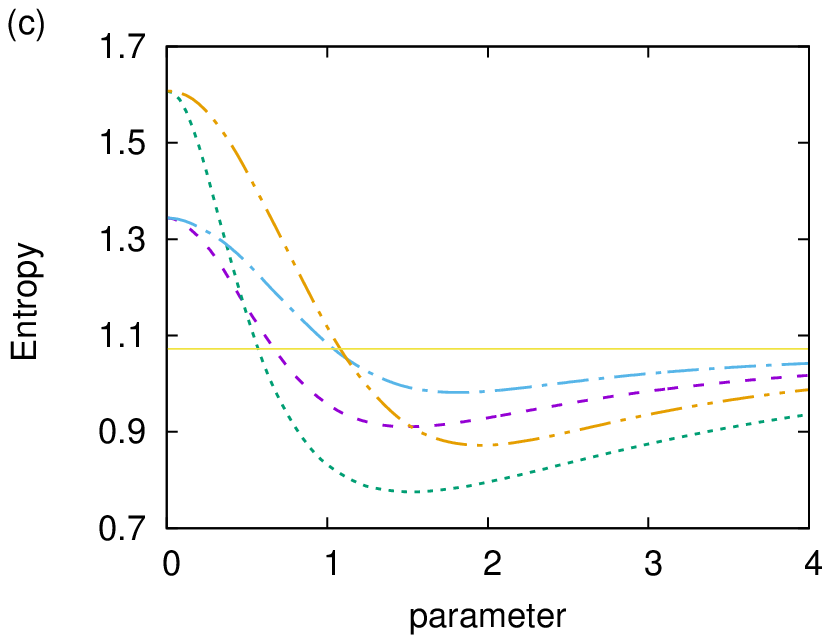}
  \caption{Tomographic entropy (a) as a function of  $\alpha$  for the ECS (violet), OCS (green) and the YS state (blue) ($\theta=\frac{1}{2}\pi$);  (b) as a function of 
  $\xi$ for the squeezed vacuum (violet) and Yuen state (green)  ($\theta=0$);  (c) 
  as a function of $\alpha$ (respectively, $\zeta)$ for $\ket{\alpha,1}$ (violet), $\ket{\alpha,3}$ (green), $\ket{\zeta,1}$ (blue) and $\ket{\zeta,3}$ (orange) ($\theta=0$). The horizontal lines denote the value below which entropic squeezing occurs.}
  \label{fig:ent_singlemode_alpha}
 \end{figure}

 \paragraph{}  We now examine the entropic squeezing  properties of single-mode states  using (\ref{eqn:single_mode_tomo_ent}). The tomographic entropy as a function of $\alpha$ for the ECS, OCS and  the YS state is squeezed in the `momentum' quadrature $P$ 
 ($\theta = \frac{1}{2}\pi$), with the OCS displaying a very different behaviour 
 than that of the ECS and the YS state (figure \ref{fig:ent_singlemode_alpha}(a)).  These three states do not display entropic squeezing  in the `position' 
 quadrature $X$  ($\theta = 0$) 
 as $\alpha$ is varied. As expected, for sufficiently large $|\alpha|$ these states show similar behaviour. In contrast,  the tomographic entropy as a function of $\xi$ for the corresponding Janus-faced partners, namely, the  squeezed vacuum and Yuen states,  exhibits squeezing in $X$  and not in $P$ (figure \ref{fig:ent_singlemode_alpha}(b)). This mirrors the phase shift of the corresponding tomograms by $\frac{1}{2}\pi$.   Further, while the ECS and its Janus-faced partner are squeezed for a wide range of parameter values, the OCS and the Yuen state are squeezed only for  $\alpha \gtrsim 0.9$ and $\xi 
 \gtrsim 0.3$.

 In contrast to these cat states, the $m$-PACS $\ket{\alpha, m} $ and the isospectral CS $\ket{\zeta, m}$  built on $\ket{m}$ display entropic squeezing in $X$ when  $\alpha$ (equivalently, $\zeta$)   is greater than a critical value  (figure \ref{fig:ent_singlemode_alpha}(c)). With increase in $m$, this critical value decreases for PACS and increases for the isospectral CS.  Further, it is clear that the critical value of $\alpha$ is higher for the  isospectral CS than for the $m$-PACS.

\begin{figure}
  \centering
  \includegraphics[scale=0.53]{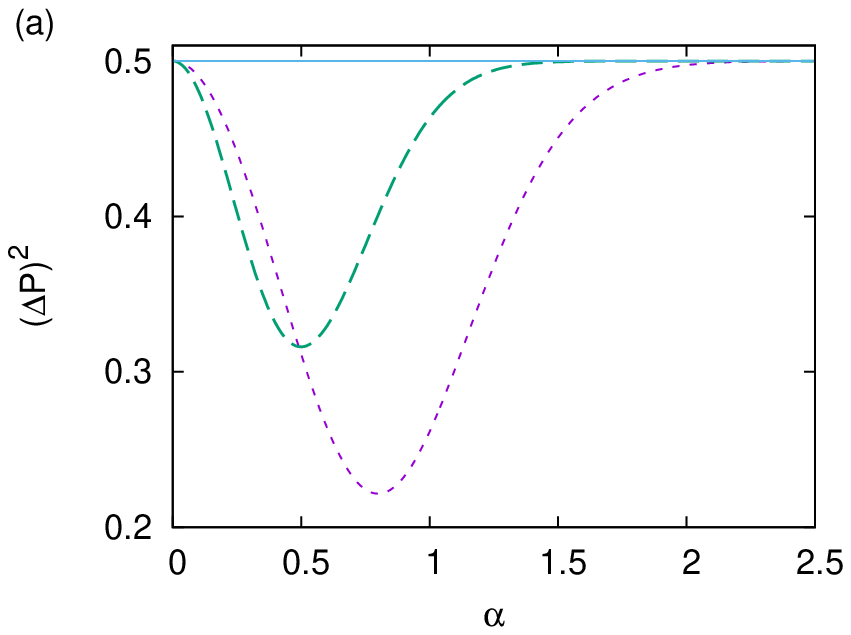}
  \includegraphics[scale=0.53]{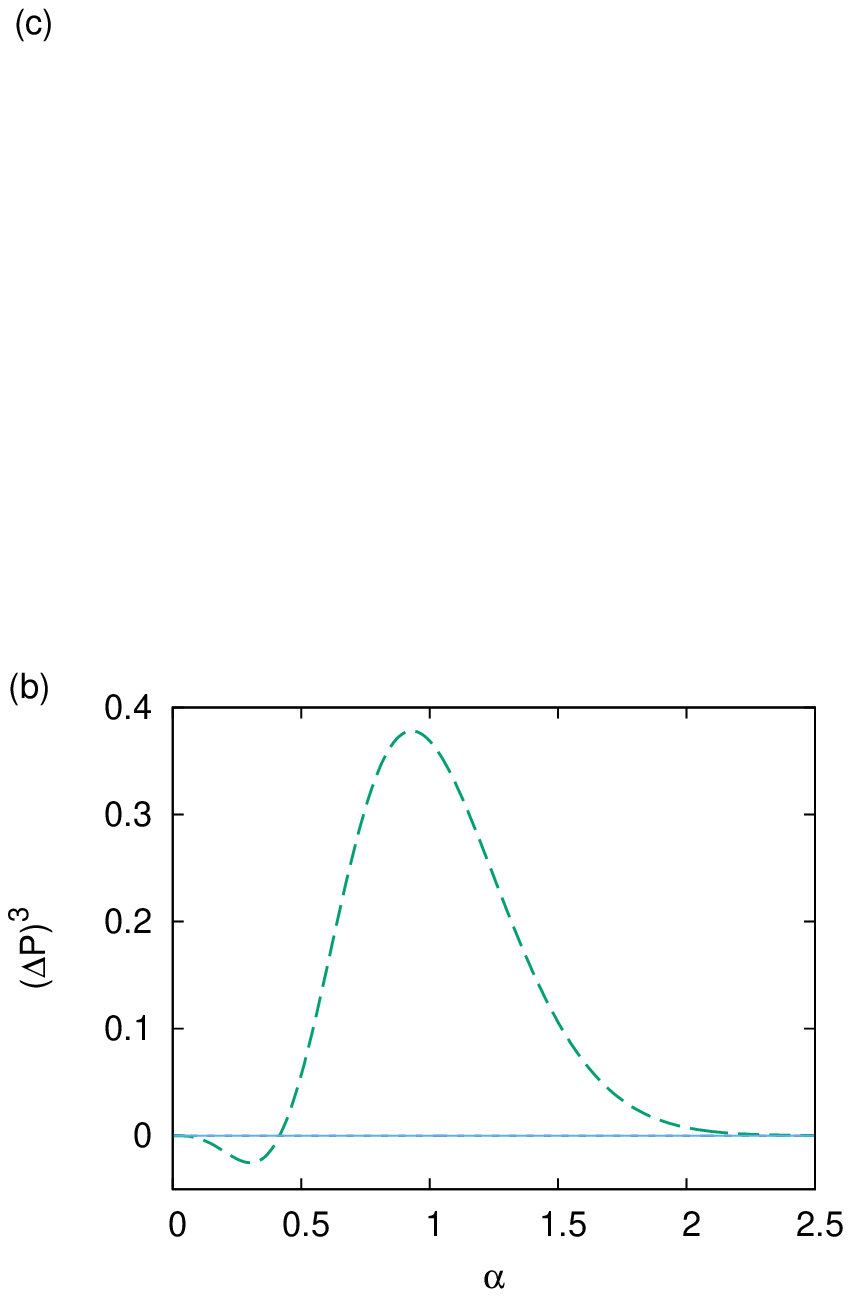}
  \includegraphics[scale=0.53]{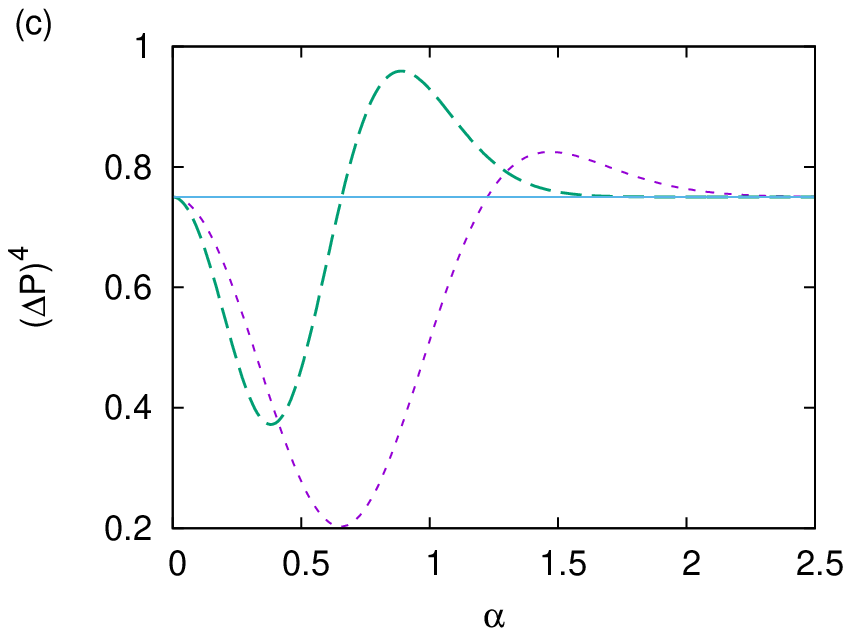}
  \caption{(a)  $(\Delta P)^{2}$, (b) $(\Delta P)^{3}$ and  (c) $(\Delta P)^{4}$ versus $\alpha$  for the ECS (violet) and the YS state (green). $(\Delta P)^{3}= 0$ for the ECS. The horizontal line denotes the value below which quadrature squeezing occurs.}
  \label{fig:var_singlemode_alpha}
 \end{figure}

 Using the procedure mentioned earlier  for calculating expectation values   
 $\aver{a^{\dagger\,k}a^{l}}$ from the tomograms,  we have verified that, like  the  entropy, the variance in $P$ corresponding to  the ECS and the YS state is squeezed  (figure \ref{fig:var_singlemode_alpha}(a)).  Similarly, it is seen that for the squeezed vacuum and Yuen state, $X$ is the squeezed  quadrature. The OCS  does not exhibit squeezing in either quadrature.  These features are consistent with the conclusion  that the extent of entropic squeezing does not reflect quadrature squeezing alone, but also includes other nonclassical effects \cite{orlowski}. As expected,  for sufficiently large $|\alpha|$,  the cat states show similar behaviour. Again, as in entropic squeezing, the $m$-PACS and the isospectral CS exhibit quadrature squeezing in $X$.

Squeezing in higher powers of $\Delta X$  and $\Delta P$ can also be quantified in a straightforward manner. A state is  squeezed to order $q$ $(q = 1,2,3,...)$ in the operator $A$ if $\aver{(\Delta A)^{q}}$  is less than the corresponding value obtained for a CS. 
$q = 2$ corresponds, of course, to the variance.
 The dependence of $(\Delta P)^{3}$ and $(\Delta P)^{4}$ on $\alpha$ is shown in figures \ref{fig:var_singlemode_alpha}(b, c). The YS state shows marginal squeezing of $(\Delta P)^{3}$  for sufficiently small values of $\alpha$ in contrast to the ECS, for which squeezing in $(\Delta P)^{3}$ is absent (figure \ref{fig:var_singlemode_alpha}(b)). 

\begin{figure}
  \centering
  \includegraphics[scale=0.78]{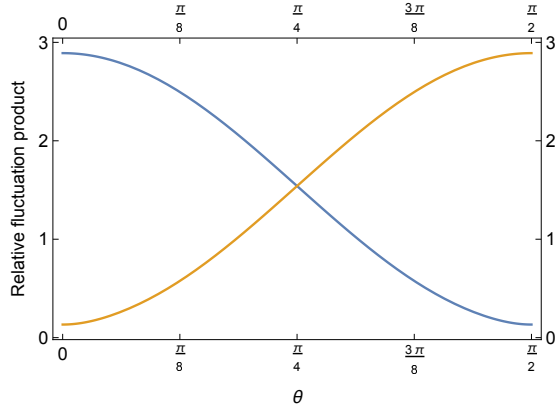}
  \caption{Relative fluctuation product between  the ECS and  the squeezed vacuum  as a function of $\theta$, for $\alpha=\xi=1$. The blue curve corresponds to $(\Delta X_{\theta})_{ECS} (\Delta X_{\theta+\pi/2})_{SQV}$, and the orange curve to $(\Delta X_{\theta})_{SQV} (\Delta X_{\theta+\pi/2})_{ECS}$.}
  \label{fig:var_singlemode_rfp}
 \end{figure}

 A signature of Janus-faced partners is exhibited in the relative fluctuations. Consider the ECS and the squeezed vacuum (SQV). It is easy to see that for fixed real values 
 of $\alpha$ and $\xi$, the product of variances $(\Delta X_{\theta})_{ECS}^{2} (\Delta X_{\theta+\pi/2})_{SQV}^{2}$ can be expressed in the form $A + B \cos\,2\theta + C \cos^{2}\,2\theta$, while  the  product $(\Delta X_{\theta})_{SQV}^{2} (\Delta X_{\theta+\pi/2})_{ECS}^{2}$ has  the form $A - B \cos\,2\theta + C \cos^{2}\,2\theta$, where $A$, $B$, and $C$ are real positive constants.  As a consequence, the square roots of these products display a symmetry property seen for instance, in figure \ref{fig:var_singlemode_rfp} where we have plotted them as a function of $\theta$,  for $\alpha = \xi = 1$. In this case, the symmetry is about the horizontal line at approximately $1.5$. A similar symmetry is seen in the relative fluctuation product corresponding to  the OCS and the Yuen state. 

 We have also verified from the relevant tomograms that the two-mode variance corresponding to the  Caves-Schumaker state exhibits squeezing for $\theta_{1}=\theta_{2} = \theta = 0$ and for the pair coherent state for $\theta = \pi/2$.  An important observation is that the two-mode tomographic entropies for these states are not squeezed, indicative of the fact that the result in \cite{orlowski} for single-mode systems (namely, that the single-mode entropy subsumes the the single-mode quadrature variance) cannot be extrapolated to bipartite systems.The reduced single-mode variances do not exhibit squeezing in any quadrature,  consistent with the fact that the reduced single-mode tomographic entropy is also not squeezed.

\section{The quantum beamsplitter: Phase dependence and squeezing properties of the output state}
 \label{sec:beamsplitter}
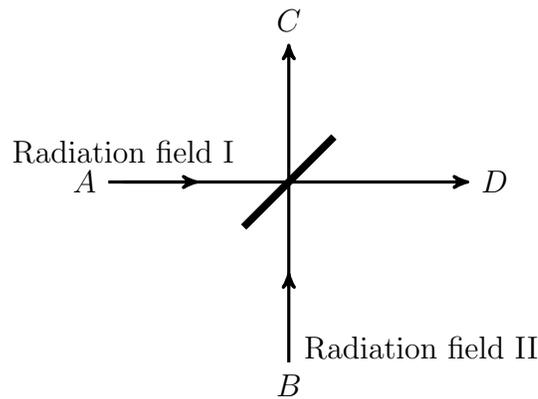
\begin{figure}[h]
 \centering
  \begin{tikzpicture}[scale=4, axis/.style={very thick, ->, >=stealth'}]
     \draw[axis] (-0.6,0) node(xline)[left] {$A$} -- (-0.3,0) ;
     \draw[axis] (-0.55,0) -- (0.6,0) node(xline)[right] {$D$} ;
     \draw[axis] (0,-0.6) node(yline)[below] {$B$} -- (0,-0.3);
     \node at (-0.55,0.1) {Radiation field I};
     \node at (0.44,-0.55) {Radiation field II};
     \draw[axis] (0,-0.35) -- (0,.46) node(yline)[above] {$C$};
     \draw[line width=1mm,] (-0.15,-0.15) -- (0.15,0.15) ;
  \end{tikzpicture}
  \caption{Schematic diagram of the quantum beamsplitter.}
  \label{fig:beamsplitter}
 \end{figure}

 In this section  we examine the squeezing properties of the output state of a 50:50  lossless beamsplitter when different superpositions of photon number states  such as  PACS or cat states are passed through the input ports 
 $A$ and $B$ (figure \ref{fig:beamsplitter}). $C$ and $D$ are the output ports. The initial states considered are  direct products of the quantum states of the radiation field. $(a, a^{\dagger}$ ), $(b, b^{\dagger})$, $(c, c^{\dagger})$ and $(d, d^{\dagger})$ are  the photon destruction and creation operators corresponding to $A, B, C$ and $D$  respectively.   The beamsplitter operation is performed by the unitary operator 
 \begin{equation}
  U = \exp\,[\textstyle{\frac{1}{4}}\pi (a^\dagger b e^{i\phi} - ab^\dagger e^{-i\phi})],
  \label{eqn:bs_op}
 \end{equation}
 where the phase difference between the reflected and transmitted fields is given by $\phi$. Hence we have 
 \begin{equation}
  c = U a\: U^{\dagger}  = (a  - e^{i\phi} b)/\sqrt{2}, \qquad  d = U b\: U^{\dagger}  = (b + e^{-i\phi} a)/\sqrt{2}.
  \label{eqn:bs_abcd_reln}
 \end{equation}

 We note that if a CS $\ket{\alpha}_{A}$ is sent through  $A$ and another CS 
 $\ket{\beta}_{B}$ through $B$,  the output state is again a direct product of two CS, $\ket{\gamma}_{C}\otimes\ket{\delta}_{D}$, where 
 \begin{equation}
  \gamma = (\alpha  - e^{+i \phi} \beta)/\sqrt{2} \quad \textrm{and} \qquad \delta =  (\beta + e^{-i \phi} \alpha)/\sqrt{2}.
  \label{eqn:bs_cs_cs}
 \end{equation}
 In contrast, for generic direct product input states the output states are entangled. We may expand these states,  for ease of numerical computation, in the photon number basis. 
 
 We first consider the case $\phi = 0$. For an input state ECS$_{A}$ $\otimes$ ECS$_{B}$ with ECS$_{A} =\mathcal{N}_{\alpha_{+}}(\ket{\alpha} + \ket{-\alpha})$ (see (\ref{eqn:ecsocs})), and ECS$_{B}$ has $\alpha$ replaced with $\beta$, the  entangled output state is
\[\mathcal{N}_{\alpha_{+}} \mathcal{N}_{\beta_{+}} \sum_{n=0}^{\infty} \sum_{m=0}^{\infty} \frac{\big(1+ (-1)^{n+m}\big)}{\sqrt{n!m!}} \big( \mathcal{C}_{+} \gamma_{+}^{n} \delta_{-}^{m} + \mathcal{C}_{-} \gamma_{-}^{n} \delta_{+}^{m} \big)  \ket{n}_{C} \otimes \ket{m}_{D}.\]
 Here,  $\mathcal{C}_{\pm} = \exp[-(|\gamma_{\pm}|^{2} +|\delta_{\mp}|^{2} )/2]$ where  $\gamma_{\pm} = (\alpha \pm e^{i\phi} \beta )/\sqrt{2}$, and $\delta_{\pm} = (e^{-i\phi}\alpha \pm \beta)/\sqrt{2}$.

 \begin{figure}
  \centering
  \includegraphics[scale=0.34]{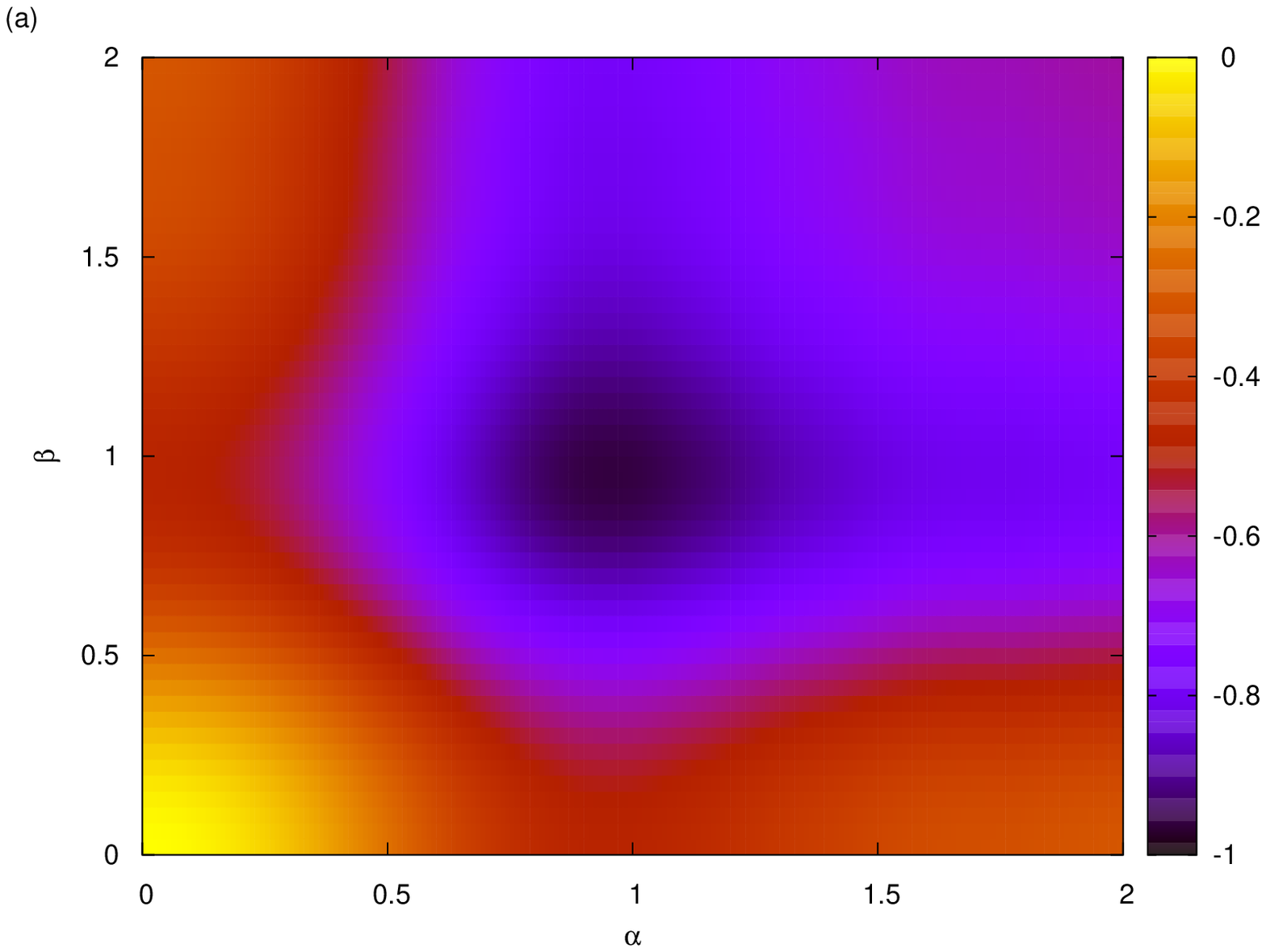}\hspace{8ex}
  \includegraphics[scale=0.34]{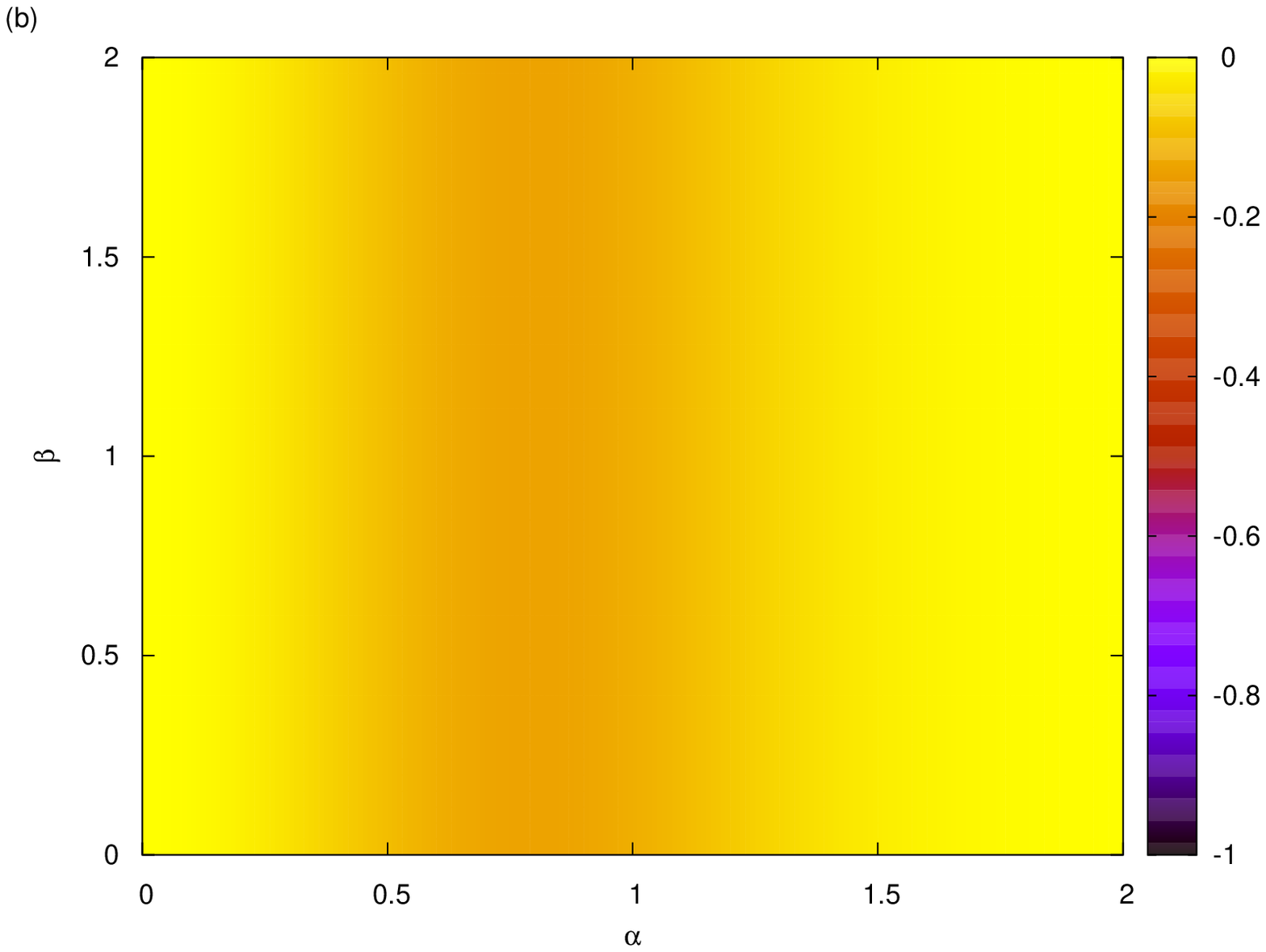}
  \caption{Contour plots for two-mode (a) tomographic entropy and (b) variance for an input state ECS$_A \otimes$ ECS$_B$ for $\phi = 0$ and  $\theta = \pi/2$. Negative values indicate squeezing.}
  \label{fig:ent_bs_ecs_ecs}
 \end{figure}
 \begin{figure}
  \centering
  \includegraphics[scale=0.34]{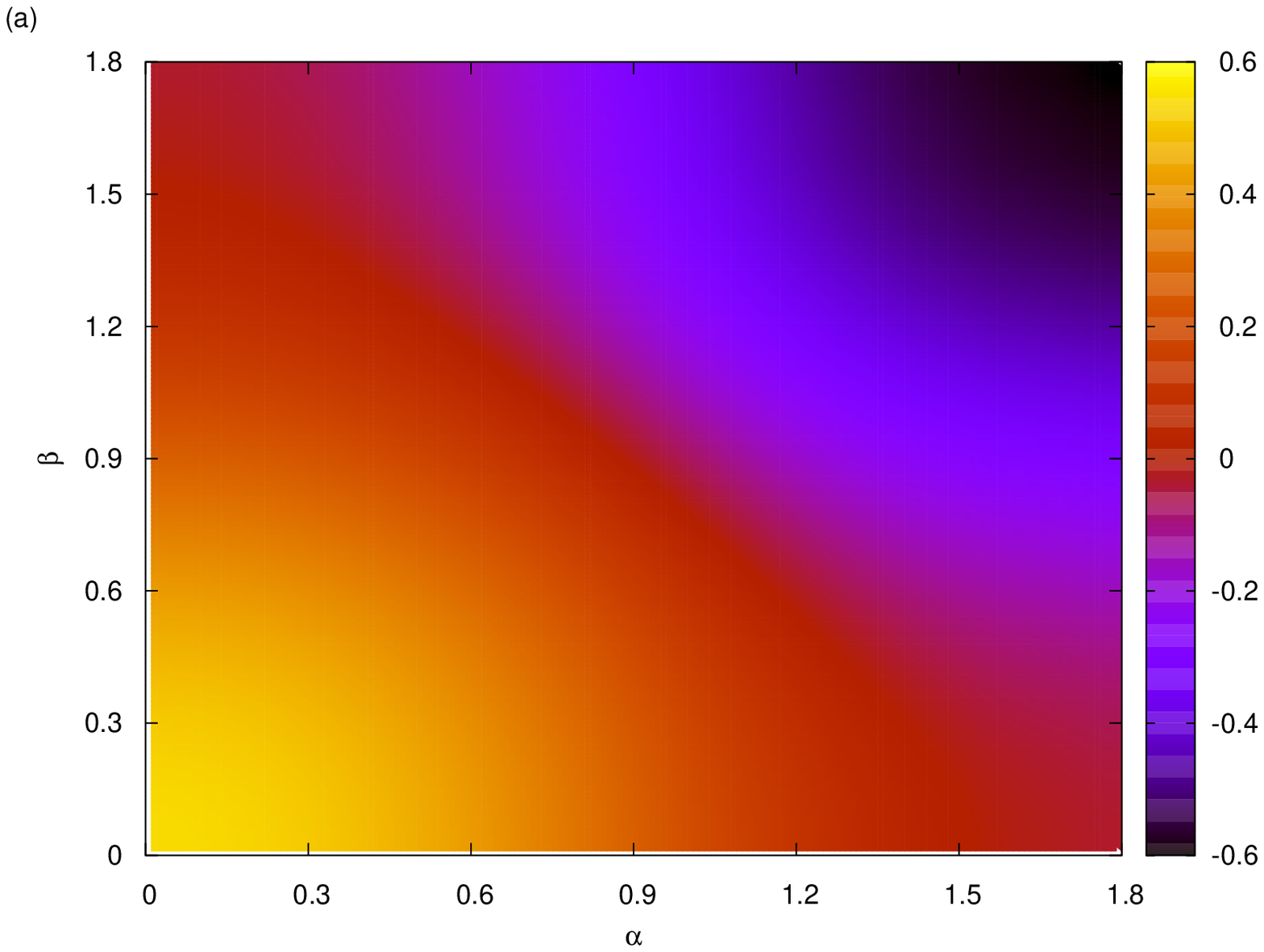}\hspace{8ex}
  \includegraphics[scale=0.34]{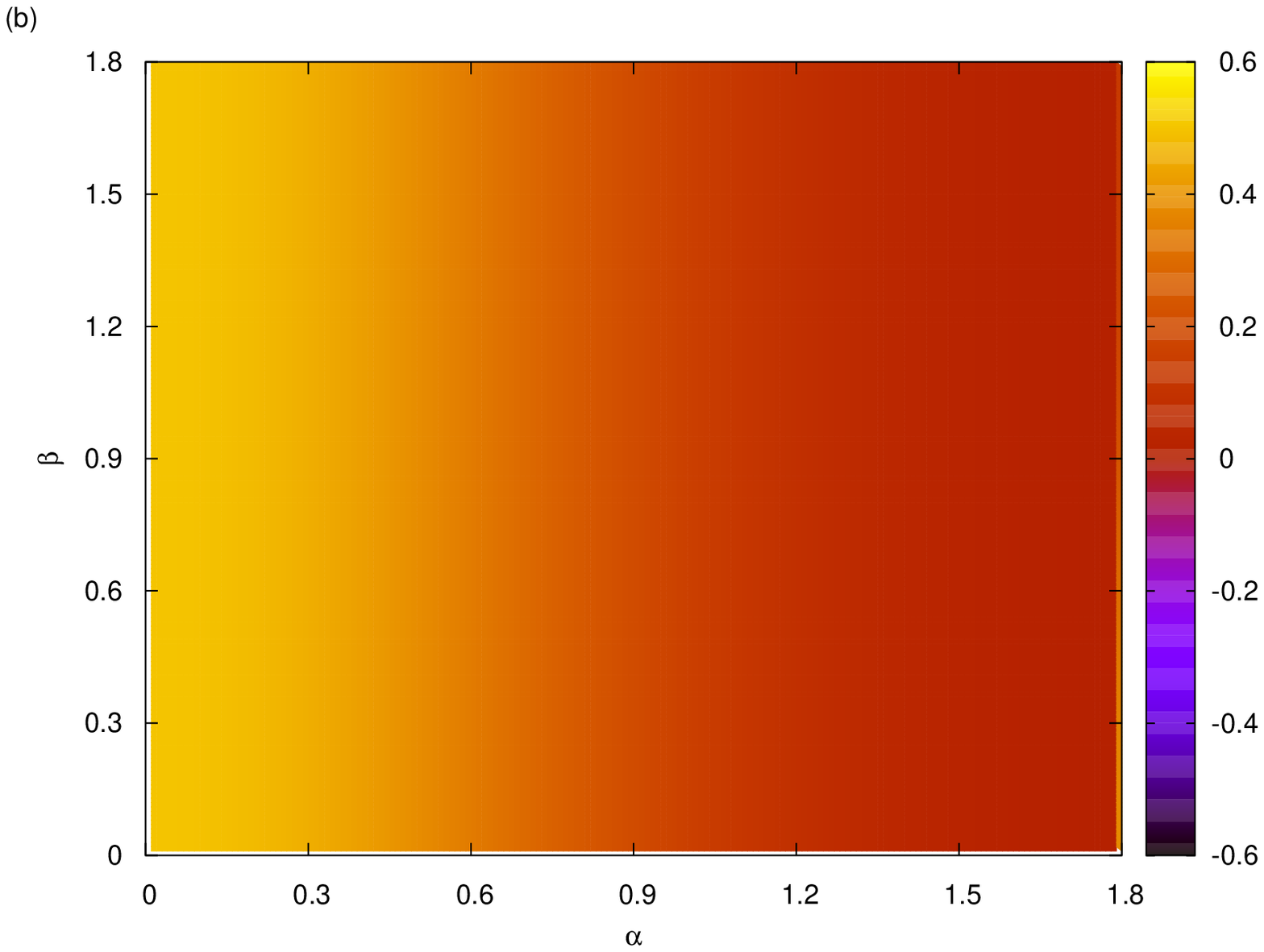}
  \caption{Contour plots for two-mode (a) tomographic entropy and (b) variance for an input state OCS$_A \otimes$ OCS$_B$ for $\phi = 0$ and  $\theta = \pi/2$.  Negative values indicate squeezing.}
  \label{fig:ent_bs_ocs_ocs}
 \end{figure}

 The contour plots (figures \ref{fig:ent_bs_ecs_ecs}(a, b)) for   both the two-mode tomographic entropy and the two-mode variance as functions of $\alpha$ and $\beta$ reveal squeezing for $\phi=0$ and $\theta = \pi/2$.  Corresponding to these values the  reduced single-mode states  also display entropic and quadrature squeezing.
  In contrast, a similar analysis for an input state OCS$_{A}\otimes$ OCS$_{B}$  (contour plots \ref{fig:ent_bs_ocs_ocs}(a, b)) reveals that only the two-mode tomographic entropy exhibits squeezing for these values of $\phi$ and $\theta$. Quadratic squeezing is absent.

 For an ECS through one port and vacuum through the other,  we have the output state
 \begin{equation}
  \ket{\Psi}_{\textrm{out}} = \mathcal{N}_{\alpha_{+}}  e^{-|\alpha|^2/2} \sum_{n=0}^{\infty} \sum_{m=0}^{\infty} \big(1 + (-1)^{n+m} \big) \frac{ \alpha^{n+m}e^{-i m  \phi}}{\sqrt{2^{n+m} n! m!}} \ket{n}_{C} \otimes \ket{m}_{D}.
  \label{eq:bs_ecs_0}
  \end{equation}

 \begin{figure}
  \centering
  \includegraphics[width=0.24\textwidth,angle=-90]{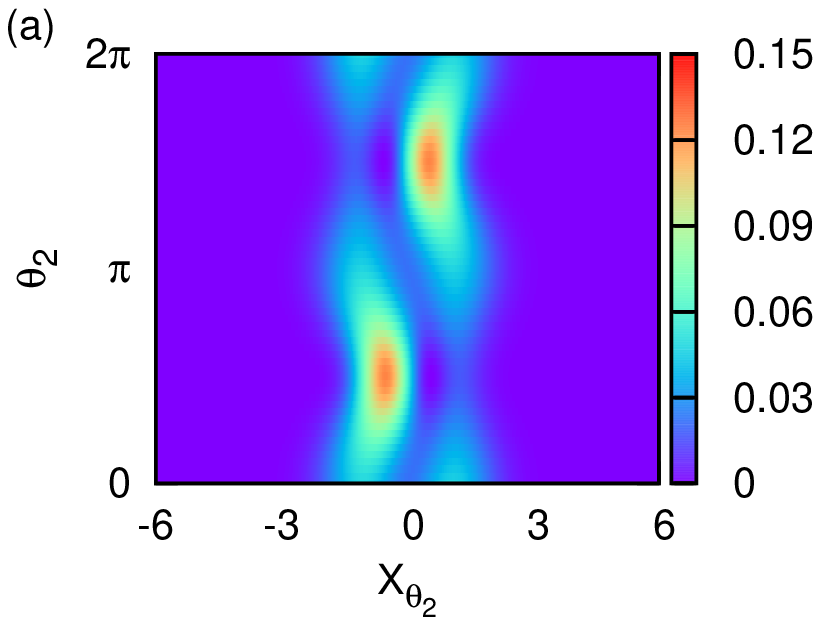}
  \includegraphics[width=0.24\textwidth,angle=-90]{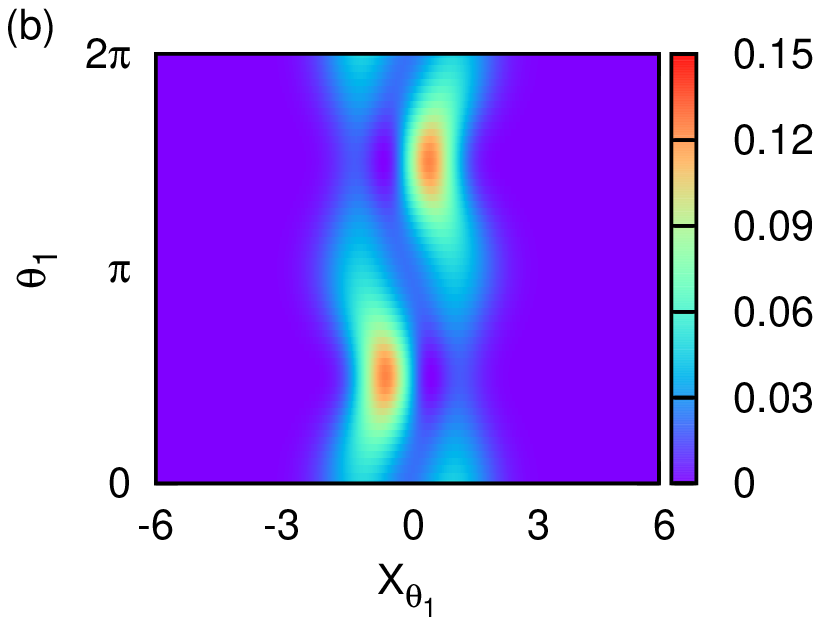}
  \includegraphics[width=0.24\textwidth,angle=-90]{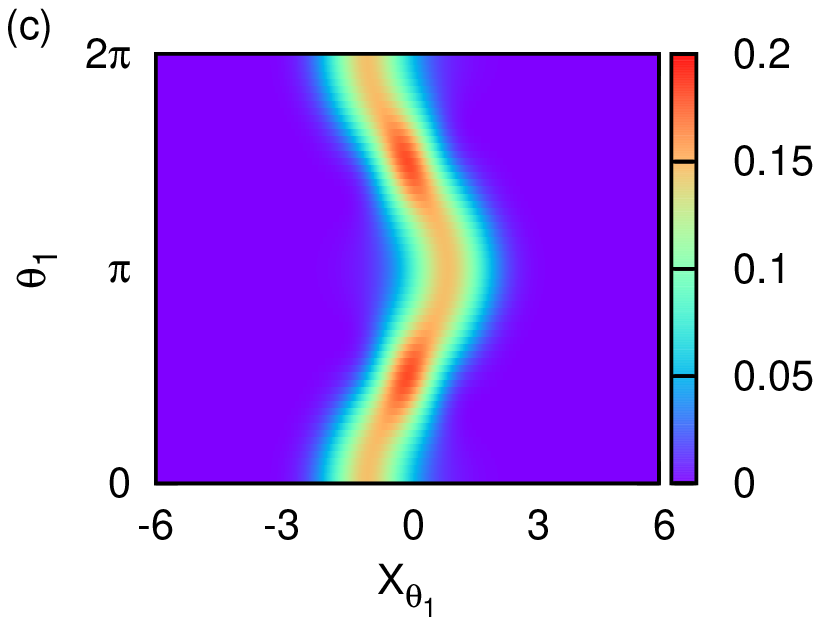}
  \caption{Output tomograms for input ECS$_A \otimes \ket{0}_{B}$ with $\alpha = 1$, (a) $X_{\theta_{1}} = 1$, $\theta_{1} = \pi/2$,  and with $X_{\theta_{2}} = 1$, $\theta_{2} = \pi/2$ for (b) $\phi = 0$ (c) $\phi = \pi/2$.}
 \label{fig:tomo1_bs_ecs_0}
 \end{figure}

 The output tomograms display interesting $\phi$ dependence.  We denote by $X_{\theta_{1}}$ and $\theta_{1}$ the tomographic variables corresponding to $C$ and by $X_{\theta_{2}}$ and $\theta_{2}$ those corresponding to $D$. The tomographic projections   obtained for different values of  $\phi$ for  a fixed value of $X_{\theta_{1}}$ and $\theta_{1}$ are qualitatively similar.  This follows from the fact that the phase dependence through $\phi$ is associated only with port $D$ (see (\ref{eq:bs_ecs_0})) and hence changes in  the tomographic variables corresponding to $C$  merely change the projection by a phase. This feature holds for any combination of unentangled input states with vacuum through one port and a cat state through the other.  The tomographic projection  \ref{fig:tomo1_bs_ecs_0}(a) is such  an  example and corresponds to an ECS through $A$ and the vacuum state through $B$.  On the other hand,  tomographic projections obtained for fixed values of $X_{\theta_{2}}$ and $\theta_{2}$  for $\phi = 0$ and $\pi/2$ are qualitatively significantly different (figures \ref{fig:tomo1_bs_ecs_0}(b, c)).
 \begin{figure}
  \centering
  \includegraphics[scale=0.55]{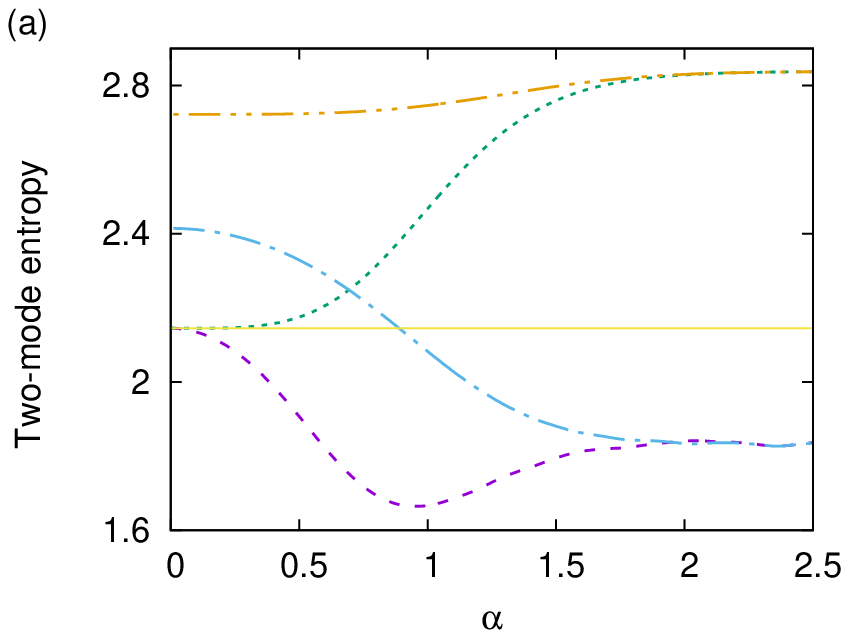}
  \includegraphics[scale=0.55]{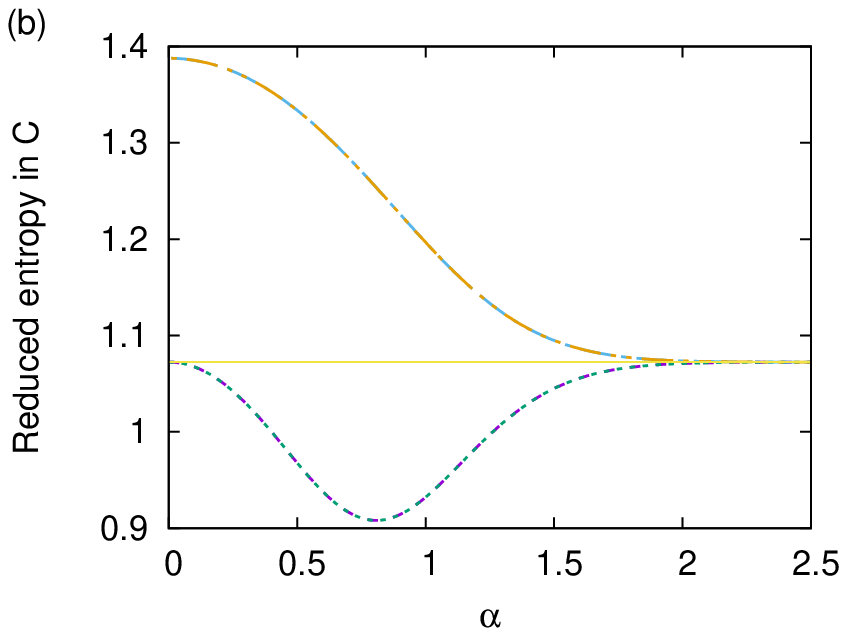}
  \includegraphics[scale=0.55]{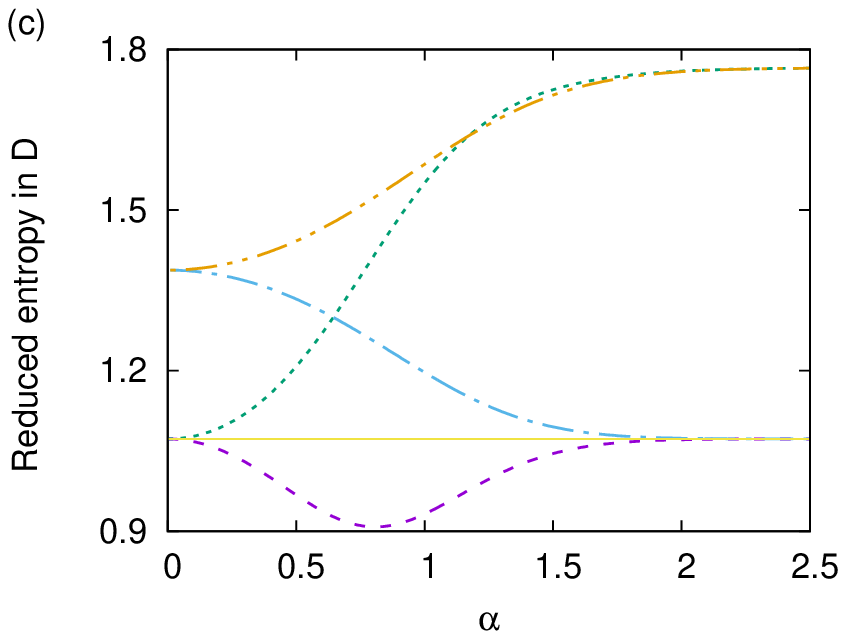}
  \caption{(a) Two-mode entropy, (b) reduced tomographic entropy in $C$ and (c)  reduced tomographic entropy in $D$ as functions of $\alpha$, for $\theta = \pi/2$ and input ECS$_A \otimes \ket{0}_B$ (violet for $\phi = 0$ and green for $\phi = \pi/2$) and OCS$_A \otimes \ket{0}_B$ (blue for $\phi = 0$ and orange for $\phi = \pi/2$).  The horizontal line indicates the value below which entropic squeezing occurs.}
  \label{fig:ent1_bs_catstate_0_alpha}
 \end{figure}
  The  variation of the two-mode  entropic squeezing and quadrature squeezing  with  $\alpha$ for $\phi = 0$ and  $\pi/2$,  setting $\theta_{1} =  \theta_{2} = \theta$, has been computed from the tomograms.  Figure \ref{fig:ent1_bs_catstate_0_alpha}(a) corresponds to a factored product  input state ECS$_{A}$ (OCS$_{A}$) $\otimes \ket{0}_{B}$ for $\theta = \pi/2$. For $\phi = 0$ the $\alpha$ dependence of two-mode entropy for these states closely resembles that of  the corresponding single-mode ECS (OCS) (compare figure \ref{fig:ent_singlemode_alpha}(a) and  \ref{fig:ent1_bs_catstate_0_alpha}(a)).  It is evident that the squeezing properties are very sensitive to $\phi$. While for both input states the outputs are squeezed over a wide range of $\alpha$ for $\phi = 0$, they are not squeezed for $\phi = \pi/2$. The beamsplitter can therefore be used with a judicious choice of values of  $\phi$ to assist  enhanced squeezing.  Quantifying squeezing directly from the tomograms is clearly a significant improvement over state reconstruction and  hence estimation of squeezing properties.
 
 The two reduced output  tomograms obtained from (\ref{eq:bs_ecs_0}) can now  be examined.  It can be inferred that one of these tomograms (corresponding to tracing out the state in port $D$) and hence the corresponding entropy  is independent of $\phi$. In this case, the variation of output state entropy with $\alpha$, for different cat states in one input port and vacuum through the other is shown in  figure \ref{fig:ent1_bs_catstate_0_alpha}(b) where we see that four curves have collapsed to two, one for each input state. Corresponding to input ECS$_{A}\otimes \ket{0}_{B}$  entropic squeezing is present. In contrast, if the reduced state is obtained by tracing out the state in $C$, $\phi$ dependence of entropic squeezing is evident from \ref{fig:ent1_bs_catstate_0_alpha}(c). The reduced output state corresponding to input OCS$_{A}\otimes \ket{0}_{B}$ does not display entropic squeezing in this case also for any value of $\phi$. 
 
  Recall that for small $\alpha$,  the tomogram for an $m$-PACS has $m$  vertical bands.  This feature is absent in the output tomogram corresponding to an input state which is a factored product of the $m$-PACS and the vacuum.

\section{Decoherence effects}
 \label{sec:decoherence}
 \paragraph{}   We investigate  how decoherence affects the output state of a beamsplitter when it interacts with a reservoir.  The reservoir is modelled by an infinite number of harmonic oscillators, initially in the ground state. We consider both amplitude decay and phase damping models.  In the former model, $\rho_{cd}(t)$ the density operator for the field state obeys the master equation \cite{gardiner}
 \begin{eqnarray}
  \frac{d \rho_{cd}(t)}{dt} = \gamma_{c} \big\{2 c \rho_{cd}(t) c^{\dagger} &-& c^{\dagger} c \rho_{cd}(t) - \rho_{cd}(t) c^{\dagger} c \big\} \nonumber \\
                                            &+& \gamma_{d} \big\{2 d \rho_{cd}(t) d^{\dagger} - d^{\dagger} d \rho_{cd}(t) - \rho_{cd}(t) d^{\dagger} d \big\}.
  \label{eqn:meq_amp}
 \end{eqnarray}
 Here $\gamma_{c}$ (respectively $\gamma_{d}$) is  the  strength of interaction between  $C$ (respectively, $D$) and the environment.  Since $\rho_{cd}(0)$ (pure output state density operator) is known, this equation can be  solved  to express $\rho_{cd}(t)$ in the photon number basis $ \ket{n}_{C} \otimes \ket{m}_{D}$ (denoted by $\ket{n;m}$). We get
 \begin{equation}
  \rho_{cd}(t) = \sum_{n,n'}\sum_{m,m'} \rho_{nn'mm'}(t) \ket{n;m}\bra{n';m'},
  \label{eqn:meq_soln}
 \end{equation}
 with
 \begin{eqnarray}
  \fl \rho_{nn'll'}(t)  =  e^{-\gamma_{n,n',l,l'}t} \sum_{r,p=0}^{\infty} \mathcal{C}_{n,n',l,l',r,p} \big(1 -  e^{- 2\gamma_{c} t} \big)^{r} \big(1 -  e^{- 2\gamma_{d} t} \big)^{p}  \rho_{(n+r)(n'+r)(l+p)(l'+p)}(0).
  \label{eqn:mel_amp}
 \end{eqnarray}
Here,  $\gamma_{n,n',l,l'} = \gamma_{c}(n + n') + \gamma_{d}(l + l')$, and $\mathcal{C}_{n,n',l,l',r,p} =  \sqrt{{n+r \choose r}{n'+r \choose r}{l+p \choose p}{l'+p \choose p}}$.
 \begin{figure}
  \centering
  \includegraphics[scale=0.4]{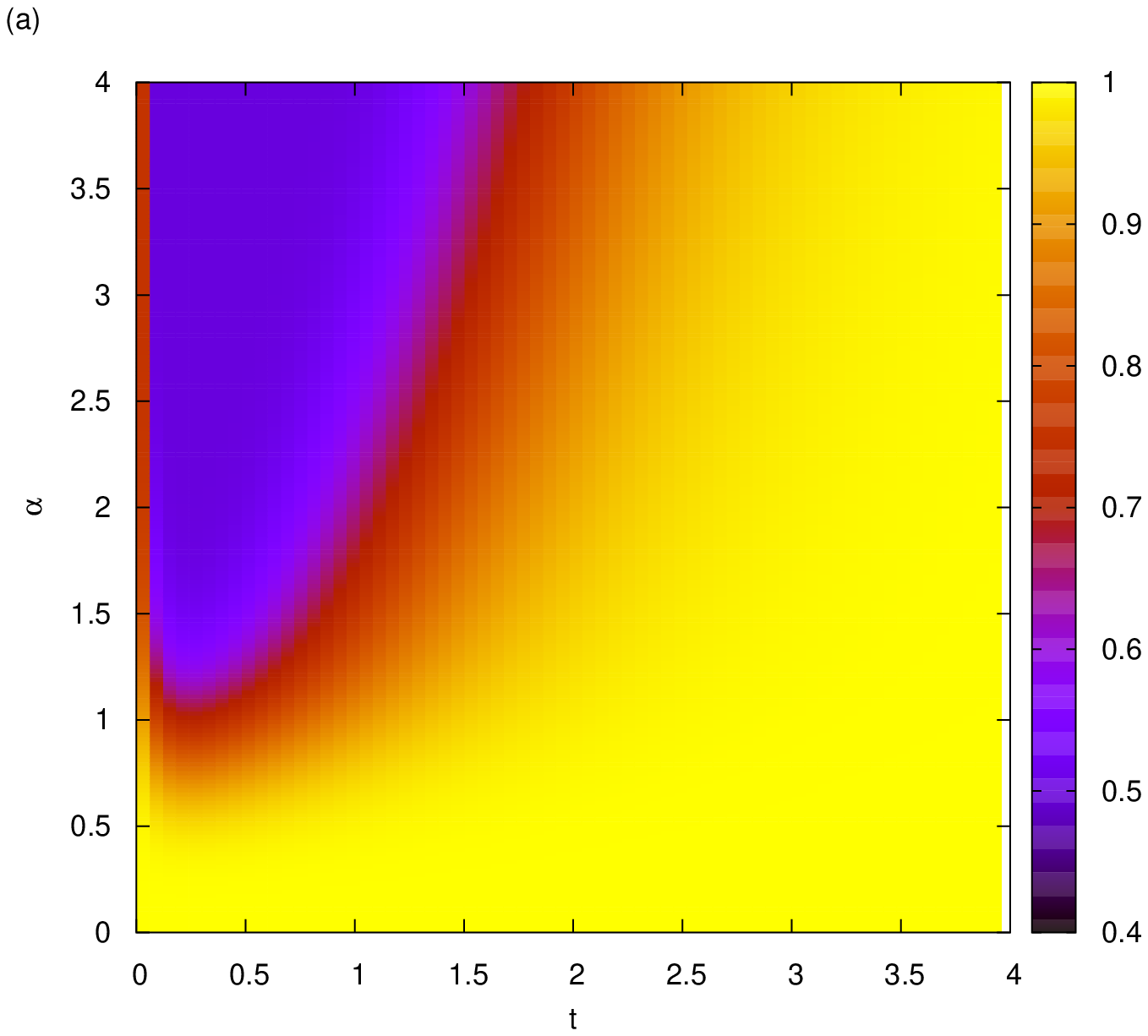}\hspace{5ex}
  \includegraphics[scale=0.42]{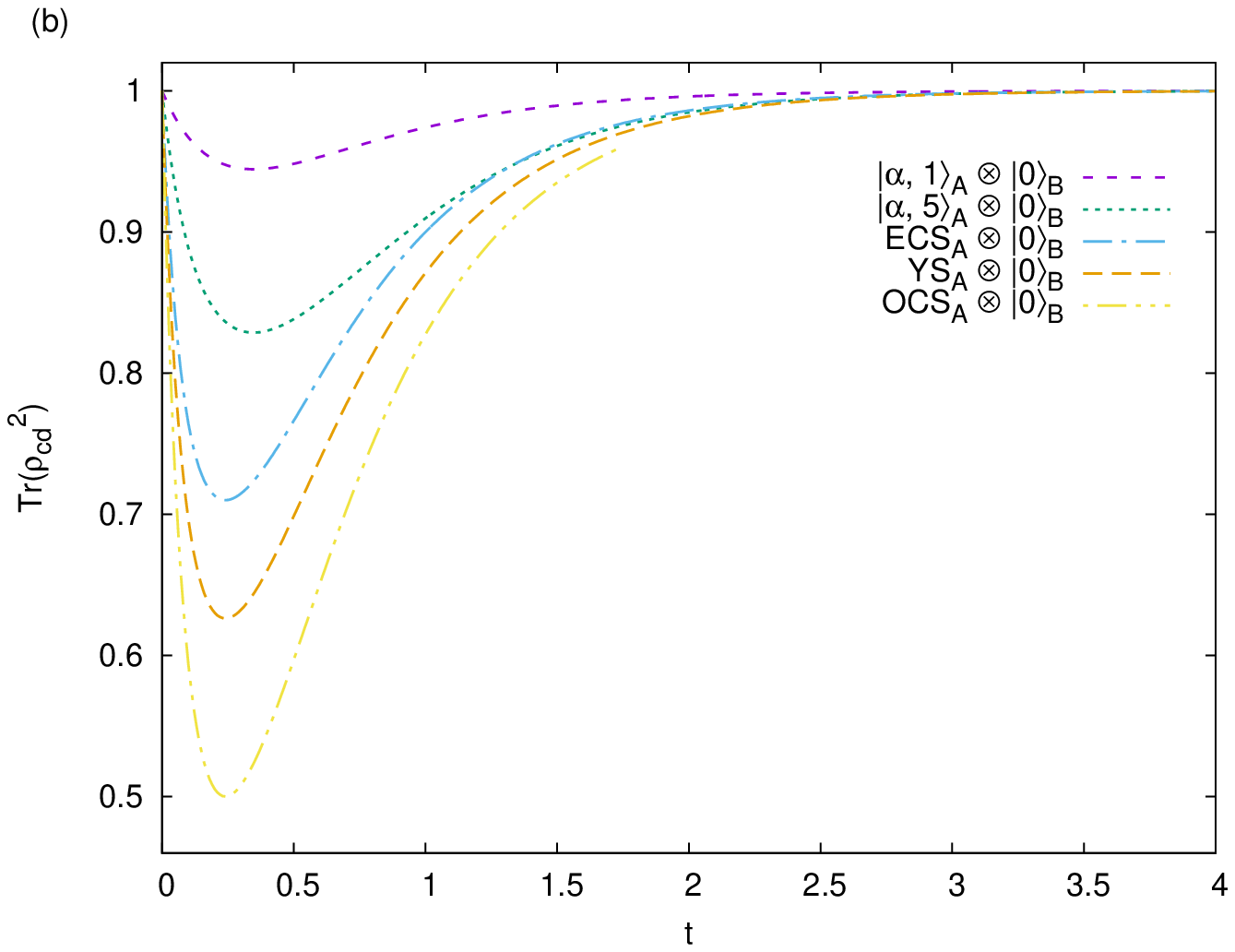}
  \caption{ Amplitude decay model: (a) contour plot of $\Tr(\rho_{cd}^2)$  as a function of time and $\alpha$ for the input state ECS$_{A} \otimes\ket{0}_{B}$ for $\phi=0$, and (b) $\Tr(\rho_{cd}^2)$  as a function of time for the input (cat state)$_{A}\otimes \ket{0}_{B}$ and $\ket{\alpha, m}_{A} \otimes \ket{0}_{B}$ ($m=1, 5$) with $\alpha = 1$, $\theta = 0$ and $\phi=0$.}
  \label{fig:decoh_amp}
 \end{figure}
 We take $\gamma_c = \gamma_d = 1$ for numerical computations.  A contour plot (figure \ref{fig:decoh_amp}(a)) reveals how $\Tr(\rho_{cd}^{2})$ depends on time and $\alpha$ for the input state ECS$_{A}\otimes \ket{0}_{B}$. We see that starting from the corresponding  output at $t = 0$ ($\Tr(\rho_{cd}^{2}) = 1$) the system evolves through a series of mixed states and after a  sufficiently long time it decoheres to  the vacuum state. Thus, the final tomogram is independent of the initial state.  As $\alpha$ increases the system takes more time to decohere completely. From figure \ref{fig:decoh_amp}(b) it is evident that for a state with marginal departure from coherence (e.g., $1$-PACS) through one input port of the beamsplitter and vacuum through the other, the  maximum departure  from unity of $\Tr(\rho_{cd}^{2}$) is substantially less than for input states with increased departure from coherence. (Compare the plots for $\ket{\alpha,1}_{A} \otimes \ket{0}_{B}$ with $\ket{\alpha,5}_{A} \otimes \ket{0}_{B}$ as inputs. Further, within  the family of cat states, the output corresponding to OCS$_{A}\otimes \ket{0}_{B}$ as input displays maximum departure from a pure state as it decoheres). It is straightforward to verify that the entropy, variance and the higher order moments (for the two-mode and both the reduced single-modes) for all these initial states also attain the value corresponding to that of the vacuum  when $\Tr(\rho_{cd}^{2})$ becomes unity at large times.
 
We now consider the phase damping model. The corresponding master equation for dissipation is
 \begin{equation}
 \fl \frac{d \rho_{cd}(t)}{dt} = \kappa_{c} (2 N_{c} \rho_{cd}(t) N_{c} - N_{c}^{2} \rho_{cd}(t) - \rho_{cd}(t) N_{c}^{2}) + \kappa_{d} (2 N_{d} \rho_{cd}(t) N_{d} - N_{d}^{2} \rho_{cd}(t) - \rho_{cd}(t) N_{d}^{2}).
  \label{eqn:meq_phase}
 \end{equation}
 Here, $N_{c} = c^{\dagger} c$, $N_{d} = d^{\dagger} d$ and $\kappa_{c}$ (respectively $\kappa_{d}$) is the coupling constant between $C$ (respectively, $D$) and the environment mode. We again expand $\rho_{cd}(t)$ in the Fock basis, analogous to (\ref{eqn:meq_soln}). In this case it can be shown that 
 \begin{equation}
  \rho_{nn'mm'}(t) = e^{ - \{ \kappa_{1}(n-n')^{2} +  \kappa_{2} (m-m')^{2}\}t } \rho_{nn'mm'}(0).
  \label{eqn:mel_phase}
 \end{equation}
 \begin{figure}
  \centering
  \includegraphics[scale=0.4]{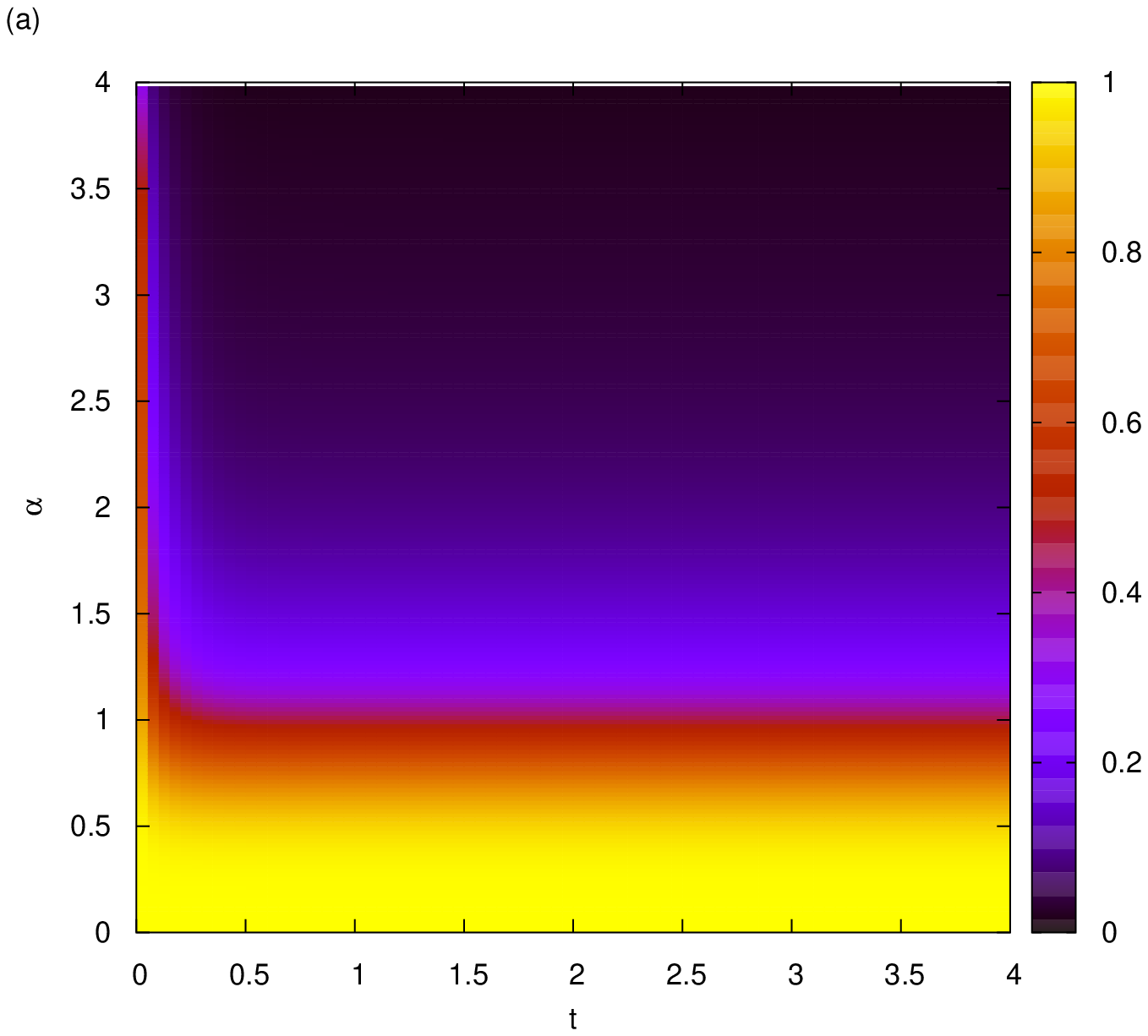}\hspace{5ex}
   \includegraphics[scale=0.42]{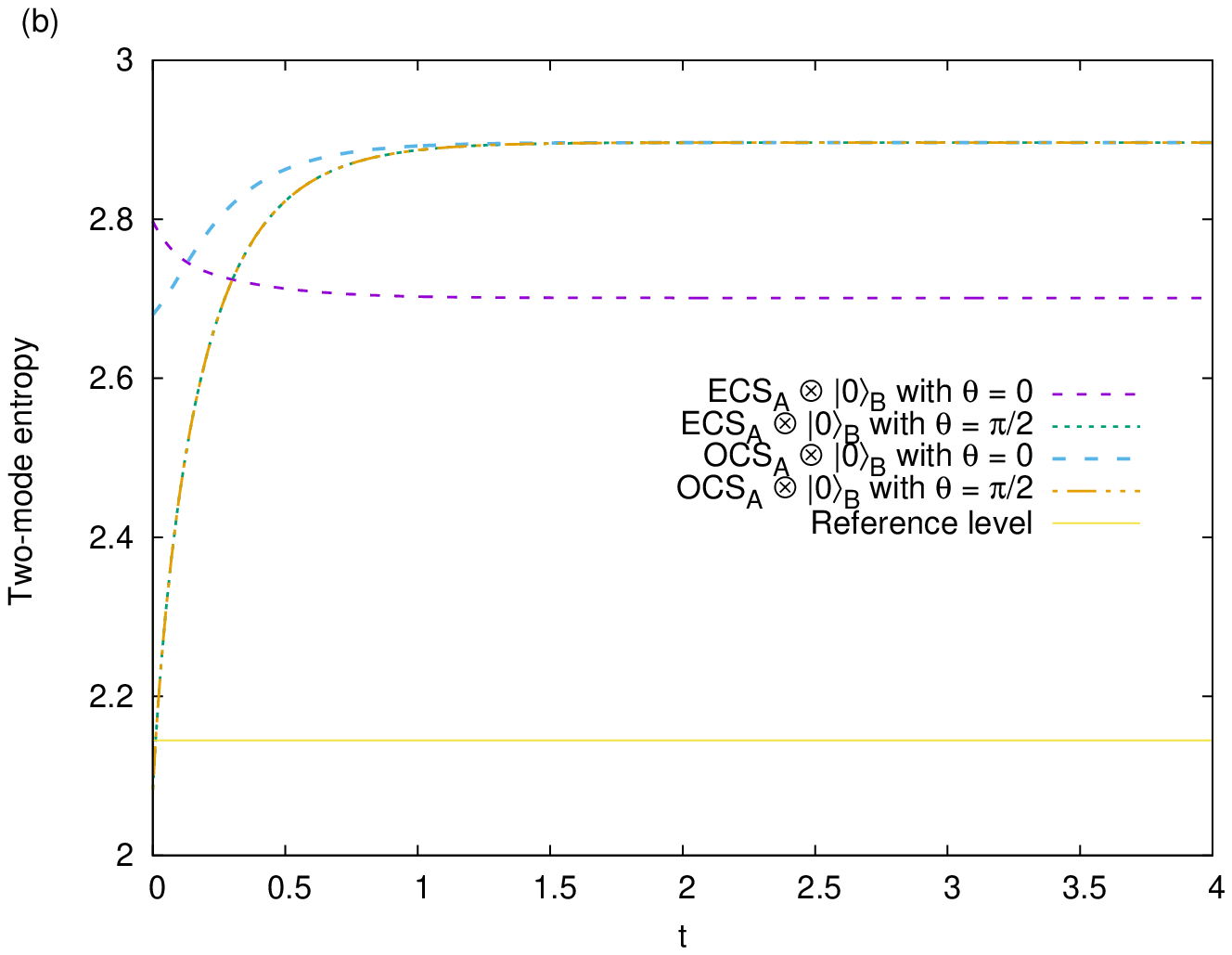}
  \caption{Phase damping model: (a) contour plot of $\Tr(\rho_{cd}^2)$  as a function of time and $\alpha$ for the input state ECS$_{A}\otimes\ket{0}_{B}$ for $\phi=0$, and (b) two-mode entropy  as a function of time for the input (cat state)$_{A}\otimes \ket{0}_{B}$ with $\alpha = 1$, $\theta = 0, \pi/2$ and $\phi=0$. The horizontal line indicates the value below which entropic squeezing occurs.}
  \label{fig:decoh_phase}
 \end{figure}
In contrast to the situation in the amplitude decay model, in this case an initial  pure state loses phase information completely after sufficiently long times. The precise form of the final mixed state (corresponding to a diagonal form of $\rho_{cd}$) depends on the initial state considered.  A contour plot  shows the variation of $\Tr(\rho_{cd}^{2})$  with time and $\alpha$ for an input ECS$_{A}\otimes \ket{0}_{B}$ (figure \ref{fig:decoh_phase}(a)). The  variation of the two-mode entropy with time corroborates this result, and for different initial states this entropy  saturates at different values (figure \ref{fig:decoh_phase}(b)).
  
 In summary: We have examined optical tomograms of several experimentally relevant states of the radiation field and identified distinctive signatures of different states in their tomograms. In particular,  we have  exploited this tomographic approach to obtain symmetry properties of  Janus-faced partner states corresponding to both single-mode and bipartite systems.  We have examined tomograms of cat states and multiphoton coherent states which have been produced in the laboratory and are ideal candidates for quantum information processing. Entropic squeezing, quadrature and higher-order squeezing properties of states have been computed directly from tomograms, thus circumventing the need for state or density matrix  reconstruction from experimentally obtained tomograms. This approach has been used to estimate the  dependence of squeezing properties of entangled states on the relative phase between the reflected and transmitted components of fields passing through a beamsplitter. Our investigation shows that  this phase can be chosen so as to 
 tailor  the  squeezing properties of various states of the radiation field---an aspect which is very important for information storage and transmission.

\appendix
\section{Expression for normal-ordered moments from optical tomograms}
 \paragraph{}  We summarise the important steps in the procedure for obtaining normal-ordered moments  for infinite-dimensional single-mode systems, from optical tomograms. The details of the calculation are  discussed in \cite{wunsche}.
 
 The density operator can be written in the normal-ordered form  
 \begin{equation}
   \rho = \sum_{k=0}^{\infty}\sum_{l=0}^{\infty} \rho_{k,l}  a^{\dagger^k}a^{l},
  \label{eqn:rho_n_ord}
 \end{equation}
 with
  \begin{equation}
  \rho_{k,l}  = \sum_{j=0}^{\{k, \:l\}} \frac{(-1)^{j}}{j! \sqrt{(k-j)! (l-j)!}} \expect{k-j}{\rho}{l-j}
  \label{eqn:rho_kl}
 \end{equation}
 in the Fock basis.  Here $\{k, l\}$ stands for $\min(k, \:l)$. 
 
 We can also express the density operator in terms of expectation values as 
  \begin{equation}
  \rho = \sum_{k, l=0}^{\infty} \ket{l}\bra{k} \Tr\,\big(\ket{k}\bra{l} \rho\big).
  \label{eqn:rho_trace}
 \end{equation}
 Since 
 \begin{equation}
  \ket{k}\bra{l} = \frac{1}{\sqrt{k! l!}} \sum_{u=0}^{\infty} \frac{(-1)^u}{u!}  a^{\dagger\,k+u}a^{l+u},
  \label{eqn:projec_op}
 \end{equation}
we get
 \begin{equation}
  \rho =  \sum_{k=0}^{\infty}\sum_{l=0}^{\infty} a_{k, l} \Tr\,\big(a^{\dagger^{k}}a^{l} \rho\big), 
  \label{eqn:rho_n_ord_tr}
 \end{equation}
where
\begin{equation}
 a_{k, l} = \sum_{j=0}^{\{k,l\}}  \frac{(-1)^{j}}{j! \sqrt{(k-j)! (l-j)!}} \ket{l-j} \bra{k-j}.
  \label{eqn:a_kl}
 \end{equation}

Recalling (\ref{eqn:xtheta_n}), namely,  
\begin{equation}
\inner{X_{\theta}, \theta}{n} = \pi^{-1/4}e^{-X_{\theta}^2/2}  e^{-i n \theta} 2^{-n/2} (n!)^{-1/2} H_{n}(X_{\theta}), 
\label{eqn:xtheta_nrecall}
\end{equation}
we obtain
\begin{equation}
  \inner{X_{\theta}, \theta}{m} \inner{n}{X_{\theta}, \theta} = \frac{e^{-X_{\theta}^2}}{\sqrt{\pi}} \frac{e^{-i(m-n)\theta}}{\sqrt{m! n!} \sqrt{2^{m+n}}} H_{m}(X_{\theta}) H_{n}(X_{\theta}).
  \label{eqn:xtheta_wunsche}
 \end{equation}
Using the identities 
$\sum_{j=0}^{k} (-1)^{j}/(j!(k-j)!) = \delta_{k,0}$ and 
 \begin{equation}
  H_{k+l}(X_{\theta}) =  \sum_{s=0}^{\{k,l\}} \frac{(-2)^{s} k! l! H_{k-s}(X_{\theta}) H_{l-s}(X_{\theta}) }{s! (k-s)! (l-s)!},
  \label{eq_}
 \end{equation}
it follows from (\ref{eqn:rho_n_ord_tr}) and (\ref{eqn:xtheta_wunsche})   that
 \begin{equation}
  \fl \omega(X_{\theta}, \theta) = \expect{X_{\theta}, \theta}{\rho}{X_{\theta}, \theta} = \frac{e^{-X_{\theta}^2}}{\sqrt{\pi}} \sum_{k, l=0}^{\infty}  \frac{e^{-i(k-l)\theta}}{\sqrt{k! l!} \sqrt{2^{k+l}}} H_{k+l}(X_{\theta}) \Tr\,(a^{\dagger\,k}a^{l}\rho)
  \label{eqn:w_xtheta_w}
 \end{equation}
Using the orthonormality property of the Hermite polynomials and the 
identity
 \begin{equation}
  \sum_{u=0}^{n} \exp(2 \pi i u j/(n+1)) = (n+1) \delta_{j,0}
  \label{eq_}
 \end{equation}
 in (\ref{eqn:w_xtheta_w}), we get
 \begin{equation}
  \fl \aver{a^{\dagger\,k}a^{l}} = C_{kl}  \sum_{m}^{k+l} \exp\,
  \bigg(\!-\frac{i(k-l)m\pi}{k+l+1}\bigg) \int_{-\infty}^{\infty} dX_{\theta} \: \omega\bigg(X_{\theta}, \frac{m\pi}{k+l+1}\bigg)  H_{k+l}(X_{\theta}),
 \end{equation}
 where $C_{kl} = k! l!/((k+l+1)!\sqrt{2^{k+l}})$. As mentioned in the text, 
 a straightforward extension of this formula holds good for bipartite systems.

\section*{References}
\bibliography{reference}

\end{document}